\documentclass[journal]{IEEEtran}
\usepackage[utf8]{inputenc}
\usepackage{graphicx}
\usepackage{xspace}
\usepackage{paralist}
\usepackage{amsmath}
\usepackage{xcolor}
\usepackage{algpseudocode}
\usepackage{algorithm}
\usepackage{url}
\usepackage{multirow}
\usepackage{tcolorbox}
\usepackage{balance}
\usepackage{verbatim}
\usepackage[english]{babel}
\usepackage{numprint}
\usepackage{authblk}

\usepackage{listings}
\usepackage{caption}
\usepackage{subcaption}
\npthousandsep{,}

\usepackage[super]{nth}

\newcommand{\gumtree}{GumTree\xspace}

\newcommand{\autotuningname}{\texttt{DAT}\xspace}
\newcommand{\fleft}{$F_{\text{left}}$\xspace}
\newcommand{\fright}{$F_{\text{right}}$\xspace}
\newcommand{\astleft}{$\mathit{AST}_{\text{left}}$\xspace}
\newcommand{\astright}{$\mathit{AST}_{\text{right}}$\xspace}

\usepackage{arydshln} 
\newcommand{\algoconfig}{algorithm configuration\xspace}
\newcommand{\algoconfigs}{algorithm configurations\xspace}

\newcommand{\totalgtconfigsforjava}{2210\xspace} 

\newcommand{\filepairsexecuted}{\numprint{9052}\xspace}

\title{
Hyperparameter Optimization for AST Differencing}

\author{Matias Martinez, \and Jean-Rémy Falleri, \and Martin Monperrus

 \IEEEcompsocitemizethanks{
 \IEEEcompsocthanksitem Matias Martinez is with  Universitat Politècnica de Catalunya, Spain. E-mail: matias.martinez@upc.edu\protect\\%

 \IEEEcompsocthanksitem Jean-Rémy Falleri is with Univ. Bordeaux, CNRS, Bordeaux INP, LaBRI, UMR 5800 and Institut Universitaire de France, France. Email: jean-remy.falleri@u-bordeaux.fr\protect\\%
 
  \IEEEcompsocthanksitem Martin Monperrus is with KTH  Royal  Institute  of  Technology, Sweden. Email: monperrus@kth.se\protect\\%
 }
  \thanks{}
}

\newcommand{\rqoptimizationimprovementdefault}{To what extent does hyperparameter optimization improve the performance of AST differencing?\xspace}

\newcommand{\rqtechnique}{
To what extent can one speed up hyperoptimization using more advanced techniques than exhaustive grid search? 
}

\newcommand{\rqlocaloptimization}{To what extent is local hyperparameter optimization effective?\xspace}

\newcommand{\rqexecutiontime}{What is the overhead of local hyperparameter optimization?\xspace}

\begin{document}

\maketitle
\thispagestyle{plain}
\pagestyle{plain}

\begin{abstract}

Computing the differences between two versions of the same program is an essential task for software development and software evolution research.
AST differencing is the most advanced way of doing so, and an active research area.
Yet, AST differencing algorithms rely on configuration parameters that may have a strong impact on their effectiveness.
In this paper, we present a novel approach named \autotuningname{} (\emph{\underline{D}iff \underline{A}uto \underline{T}uning}) for hyperparameter optimization of AST differencing.
We thoroughly state the problem of hyper-configuration for AST differencing.
We evaluate our data-driven approach \autotuningname{} to optimize the edit-scripts generated by the state-of-the-art AST differencing algorithm named GumTree in different scenarios.
\autotuningname{} is able to find a new configuration for GumTree that improves the edit-scripts in 21.8\% of the evaluated cases.

\end{abstract}

\section{Introduction}
\label{sec:intro}

The computation of the differences between two versions of the same program is an essential task for software development.
It is done on a daily basis when developers share and discuss code changes, through pull-requests in development platforms (eg. Github, Gitlab, etc) or through patches over mailing lists (patches on the Linux Kernel Mailing List).
The most common differencing strategy of developer tools is line-based: 
a diff is a list of \emph{chunks} where each chunk is a set of added and/or removed lines. For example, \texttt{git} provides the widely used command `git-diff` for computing a line-based diff script.
To push this state-of-practice further, there is an active line of research on differencing algorithms that work at the level of Abstract Syntax Trees (AST) instead of lines \cite{Chawathe1996, fluri_change_2007, falleri_fine-grained_2014, huang_cldiff_2018,dotzler_move-optimized_2016,frick_generating_2018}.
In this case, the differences between two program versions are expressed in terms of actions over nodes from the trees under comparison (e.g., Insert, Remove, Update and Move nodes). 
AST differencing has been shown to be significantly better than line differencing for a number of tasks~\cite{falleri_fine-grained_2014, decker_srcdiff_2020}. 

AST differencing is of great importance not only for practitioners, but also for software engineering research.
AST differencing algorithms have been extensively used to study software evolution \cite{kim2011evolutionapi, Fluri2007Evolution, Tsantalis2018Refactor}.
Key usages of AST differencing are pattern inference for automated program repair \cite{le2017s3, fixminer,Le2016HDRepair}, bug fix analysis \cite{Zhong2015Fixes, Sobreira2018, Martinez2013Models}, code recommendation \cite{Meng2013Lase, Meng2011Sysedit, Nguyen2016}.
Consequently, it is of great importance for both practitioners and researchers to have reliable AST differencing tools.

Virtually all AST differencing algorithms can be configured in some way. 
That is, they have \emph{hyperparameters} that control and guide the differencing process. 
For example, the state-of-the-art differencing tool GumTree has one hyperparameter called \texttt{BUM\_SMT} for setting a minimum similarity threshold between two trees. 
This parameter has a default value of $0.5$. 
Changing that value impacts on the produced edit-scripts.

Unfortunately, as shown in previous research~\cite{fan2021differential,delaTorre2018Imprecision}, state-of-the-art tools such as GumTree often generate sub-optimal diffs, that is, edit-scripts that contain \emph{spurious} AST operations.
\emph{In this paper, we show that these hyperparameters matter and that one can improve AST differencing performance by finding hyperoptimized values.}

In this paper, we present a novel approach named \autotuningname (\emph{\underline{D}iff \underline{A}uto \underline{T}uning}) for hyperparameter optimization of the AST differencing.
Given an AST differencing algorithm, 
\autotuningname{} finds the optimal parameter configuration given a set of file-pairs to be compared.
\autotuningname{} requires neither ground-truth nor labeled file pairs, the optimization is guided by a widely accepted quality metric for AST differencing~\cite{falleri_fine-grained_2014, higo_generating_2017, hashimoto_diffts_2008, huang_cldiff_2018, frick_generating_2018, dotzler_move-optimized_2016, matsumoto_beyond_2019}: the length of the edit-script.
To our knowledge, we are the first to thoroughly study the problem of suboptimal configuration for AST differencing and to propose a data-driven approach to solve this problem.

We evaluate \autotuningname{} by searching for the best configurations of the AST differencing algorithm GumTree. 
In particular, we evaluate it in two scenarios.
The first scenario consists of searching for the configuration that works best on a set of file pairs.
In this case, \autotuningname{} does \emph{global hyperparameter optimization.}
The configuration found can be used as the new default configuration when a practitioner aims to apply diffs in a new programming language or AST meta-model.
Secondly, we evaluate the ability of \autotuningname{} to find the best configuration for a particular case (i.e., a single file-pair).
In this case, \autotuningname{} does \emph{local hyperparameter optimization}.
Our evaluation consists of executing those optimizations on more than \numprint{9000} pairs of Java files extracted from real-world commits of open-source Java projects.

The research question that we invsetigate in this paper are:
\begin{itemize}
\item {RQ1: \rqoptimizationimprovementdefault}
\item {RQ2: \rqtechnique}
\item {RQ3: \rqlocaloptimization}
\item{RQ4: \rqexecutiontime}
\end{itemize}

Our experimental results show that 
\begin{inparaenum}[\it 1)]
\item \autotuningname{}  is able to find a hyperoptimized configuration which produces shorter edit-scripts for 21.8\% of the cases using the JDT meta-model (which is the meta-model used by default by GumTree).
\item  Local hyperparameter optimization is effective, as it allows finding shorter edit-scripts than the default configuration in up to $\approx$26\% of cases.

\end{inparaenum}

To sum up, the contributions of this paper are:

\begin{itemize}
\item An original data-driven approach, called \autotuningname for hyperparameter optimization in the context of AST differencing.
\item An original and sound protocol for studying the performance of AST differencing, with cross-validation and statistical validation. The results of this experiment are available at \url{https://github.com/martinezmatias/dat-experimental-results}.
\item A publicly available tool for hyper-optimization of AST differencing. It natively supports the popular GumTree AST differencing engine and provides extension points to integrate other differencing tools:
\url{https://github.com/martinezmatias/diff-auto-tuning}.
\end{itemize}

The paper continues as follows.
Section \ref{sec:terminology} presents the terminology used in the paper.
Section \ref{sec:motivation} presents three cases in which state-of-the-art AST differencing produces incorrect results.
Section \ref{sec:hyperparameter} discusses hyper-parameters of differencing algorithms.
Section \ref{sec:dat} presents \autotuningname{}, our hyperparameter optimization approach for AST differencing.
Section \ref{sec:results} presents the results of the evaluation.
Section \ref{sec:threatstovalidity} presents the threats to validity.
Section \ref{sec:relatedwork} presents the related work.
Section \ref{sec:conclusion} concludes the paper.

\section{Terminology}
\label{sec:terminology}
In this section, we present the key terminology related to our contributions.

\paragraph*{AST Differencing Algorithm} computes the differences between two ASTs (Abstract Syntax Tree).
For example, ChangeDistiller \cite{fluri_change_2007} and GumTree \cite{falleri_fine-grained_2014} are two popular AST diff algorithms.  
Typically, an AST diff algorithm has two input parameters, the two ASTs (\astleft and \astright) generated from two source code files (\fleft and \fright, respectively).
Those ASTs are modeled using a \emph{meta-model} (defined below).
Finally, the AST algorithm outputs the differences between the two ASTs in the form of an \emph{edit-script} (defined below).
An AST differencing algorithm is implemented in a tool, for example, 
GumTree \cite{falleri_fine-grained_2014} is implemented in GumTreeDiff.\footnote{\url{https://github.com/GumTreeDiff/gumtree}}

\paragraph*{Edit-script} is the output of an AST diff algorithm.
It represents the result of the comparison between the two ASTs given as input parameters. 
The edit-script is a sequence of edit operations applied to the AST nodes of \astleft.
The operations are typically typed as \{insert, remove, update, move\}, and are meant to represent the transformation from \astleft into \astright.

\paragraph*{Matcher} 
For computing an edit-script between two ASTs, \astleft and \astright, it is necessary to try to match (i.e., \emph{link}) some tree nodes of \astleft with nodes from \astright.
The algorithm that executes this task is known as \emph{matcher}.
The criterion for matching two nodes depends on the matching algorithm, and can consider, for instance, the node's type and label, the topology, etc.
The list of matched and unmatched nodes is then used for deducing an edit-script. 
For instance, those nodes from \astleft and \astright that could not be matched correspond to removed or inserted nodes, respectively.

\paragraph*{AST Meta-model} defines the structure of an AST.
In particular, the meta-model describes:
\begin{inparaenum}[\it a)]
\item the possible types of AST nodes (e.g, invocations, assignments, methods);
\item the attributes of each node type (e.g., a label);
\item the optional and mandatory children of each node type, if any.
\end{inparaenum}

For example, a simplified meta-model for the Java language may have 4 node types for \emph{classes}, \emph{methods}, \emph{fields} and \emph{statements}; the \emph{method} nodes have
\begin{inparaenum}[\it a)]
\item an attribute called \emph{name}, 
\item a list of zero or more \emph{statements} nodes as children.
\end{inparaenum}

Note that there can be several meta-models for representing ASTs of the same programming language. 
For example, \gumtree can model a Java AST according to four different meta-models: JDT, Spoon, JavaParser and srcML.
Since the choice of meta-model has an impact on the topology of the resulting ASTs, it finally also has an impact on the computed edit-scripts.

\section{Motivation}
\label{sec:motivation}

Previous work has revealed that state-of-the-art AST differencing algorithms generate inaccurate mappings, which impacts on the quality of the edit-scripts  generated.
For example, Fan et al.~\cite{fan2021differential} show that GumTree~\cite{falleri_fine-grained_2014} generates inaccurate mappings for 20\%-29\% of the file pairs analyzed.

In this Section, we present two cases for which GumTree with its default configuration produces incorrect or non-optimal edit-scripts.
Then, in Section \ref{sec:results}, we show how autotuning GumTree using our approach \autotuningname{}  allows GumTree to produce more understandable edit-scripts for these two cases.

\begin{figure}[t]
\caption{(Case 1) Example of spurious add-remove edits found by GumTree using the default configuration and JDT meta-model.
Other no-spurious edits are not presented in the figure.}
\includegraphics[width=\columnwidth]{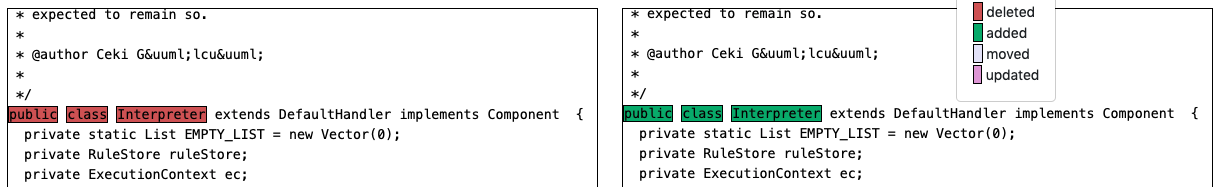}

\centering
\label{fig:case:spurious}
\end{figure}

\subsection{Case 1: Spurious Add-Remove}
\label{sec:motivation:case1}

Diff algorithms such as GumTree can produce edit-scripts with spurious edits~\cite{delaTorre2018Imprecision}.
A recurrent case of these edits is presented in Figure \ref{fig:case:spurious}, which shows the \texttt{Interpreter.java file} from the Log4J project. It presents the version from commit a04d92 (right part) and its previous version (left part).
The figure also shows in a colored box the code affected by the edits obtained by GumTree after computing the diff of those two files, using the default configuration and the JDT metamodel.
Each color represents a different edit type (red corresponds to remove, green to insert). 
The edit-script includes, in total, six edits. 
Three edits remove tree tokens (\texttt{public}, \texttt{class} and \texttt{Interpreter}), each one represented by an AST node (left part, in red), and the other three edits insert the same three tokes (left part, in green).
All these are spurious edits. 
GumTree should not generate any of these six edits.
As we discuss later in this paper, GumTree using the best configuration found by \autotuningname does not produce those edits.

\subsection{Case 2: Including Updates in the Edit-script}
\label{sec:motivation:update}

\begin{figure}[t]
\centering
\caption{(Case 2) Visualization of the edits computed by GumTree using default configuration.
}
\includegraphics[width=\columnwidth]{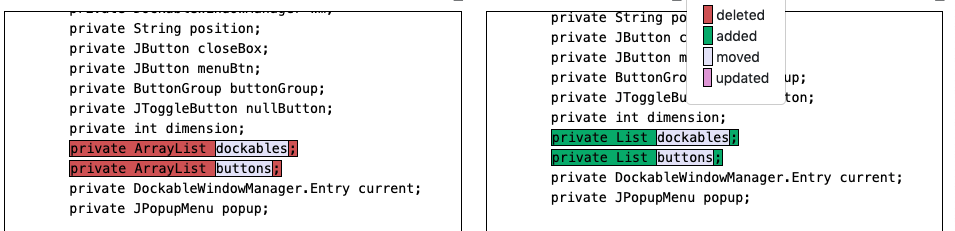}
\label{fig:case:update:default}
\end{figure}

We now focus on the differences that GumTree detects between the version of the file \texttt{PanelWindowContainer.java} from the jEdit project in commit 6867bd and its previous version.

The changes made by the developer were two: updates of the type of variables \texttt{dockables} and \texttt{buttons} from \texttt{ArrayList} to \texttt{List}.
However, the diff produced by GumTree looks like that produced by a line-based diff such as GNU diff:
The edit-script removes the fields from the left part, adds the fields from the right part (using the new field's type \texttt{List}), and moves the tokens related to the field names.
This edit strip does not clearly express the changes done by the developers (i.e., update two nodes), and introduces spurious edits (e.g., the remove and add of tokens `private').

\subsection{Fixing cases with spurious edits}

AST differencing algorithms detecting move edits (such as GumTree) are all based on heuristics because this problem is NP-hard. The downside is that node mapping heuristics can behave badly and often make use of hyperparameters.  In this paper, our goal is to optimize hyperparameters for improving the quality of the edit-scripts.
Similar improvements may eventually also be achieved by adding new features to the algorithm or proposing new algorithms.
Nevertheless, even the improved algorithms are likely to use hyperparameters, so even them can leverage our work on hyper-optimization.

\section{Hyper-parameters of Differencing Algorithm}
\label{sec:hyperparameter}

\begin{table*}[t]
\centering
\def\arraystretch{1.1}
\caption{Hyperparameter space for the GumTree AST Differencing Algorithm.}
\begin{tabular}{|l|p{8cm}|r||r|r|r||r|}
\hline
\multirow{2}{*}{Hyperparameter} & \multirow{2}{*}{Description}& \multirow{2}{*}{Default}&
\multicolumn{4}{c|}{Values}\\ 
\cline{4-7}
&&&Min&Max&Step&Total\\

\hline
\hline
Bottom-up Matcher & Bottom-up matcher used to compute the diff &Classic&\multicolumn{3}{|c||}{\{Classic, Simple, Hybrid\}}&3\\
\hline
STM\_PC&Indicates the priority calculator used by the subtree matchers&Height& \multicolumn{3}{|c||}{\{Size, Height\}}&2\\
\hline
STM\_MPTH&Threshold on the minimum priority value computed using STM\_PC &1&1&5&1&5\\
\hline
BUM\_SMT&Threshold on the minimum similarity between two AST nodes&0.5&0.1&1&0.1&10\\
\hline
BUM\_SZT&Threshold on the maximum size of AST nodes to match&1000&100&2000&100& 20\\
\hline

\hline
\end{tabular}
\label{tab:hyperparameters}
\end{table*} In this section, we describe the hyper-parameters of GumTree~\cite{falleri_fine-grained_2014}, one of the most used AST differencing algorithms.
It has been used in hundreds of research works relying on AST differencing, according to the citations of the GumTree paper \cite{falleri_fine-grained_2014} collected by Google Scholar.

\subsubsection{Hyperparameters}

First, we specify the hyperparameter space of GumTree. This is done by analyzing the source code and discussing it with the lead developers of GumTree, one of whom is also coauthor of this paper. Table~\ref{tab:hyperparameters} lists and explains the GumTree hyperparameters.
For each hyperparameter, the column \texttt{Default} shows the default value used in the implementation in version 2.1.2.\footnote{GumTree version considered:~\url{https://github.com/GumTreeDiff/gumtree/commit/ed3beeab1e00a31f23ab5e9a8292c3168221a1ca} (July 2020).}

\paragraph{Bottom-up Matcher}
\label{sec:gumtree:matcher}

The first hyperparameter of GumTree is the \emph{bottom-up matching algorithm}.
GumTree sequentially applies two types of matchers~\cite{falleri_fine-grained_2014}:
\begin{enumerate}[\it 1)]
\item Top-down matchers (also known as subtree matcher):
find isomorphic subtrees of decreasing height or size. Mappings are established between the nodes of these isomorphic sub-trees.
They are called \emph{anchors} mappings.
\item Bottom-up matchers: navigate a tree in post-order (e.g., visit first leaves, then their parents, etc.) in order to match nodes not previously matched in the top-down phase.
Two nodes match if their descendants (children of the nodes) include a large number of common anchors. Whenever a new mapping is established during this phase, a \textit{recovery} phase is applied as the last chance to find mappings of the descendants of the nodes.
\end{enumerate}

There is only one stable top-down matcher in GumTree, while there are three different stable bottom-up matchers: classic, simple, and hybrid. These three matchers differ only in the way they apply the recovery phase.

\paragraph{Priority calculator}
\label{sec:hyper:prioritycalc}
During top-down matching, GumTree greedily matches whole isomorphic subtrees. To establish the priority of the chosen subtrees, it uses a metric based upon the topology of the subtree: either its size (number of nodes in the subtree) or its height (length of the longest path from one leave to the root of the subtree). For example, when using size, GumTree will first try to find an isomorphic subtree for the subtrees with the largest number of nodes. This hyperparameter is called \texttt{STM\_PC}.

\paragraph{Minimum priority threshold}
As explained in the previous paragraph, GumTree uses a topological metric to order the subtrees to be matched by the top-down matcher. This hyperparameter indicates the minimum value of size or height (according to \texttt{STM\_PC}) to be considered by the matcher.
This hyperparameter is called \texttt{STM\_MPTH}.
The effect of the value depends on the chosen priority calculator. 
For example, GumTree configured with \texttt{STM\_PC} = size and \texttt{STM\_MPTH} = 3  will not consider subtrees with two nodes or less during top-down matching. 

\paragraph{Minimum similarity threshold}
\label{sec:methodology:mst}
A bottom-up matcher matches two AST nodes if 
\begin{inparaenum}[\it 1)]
\item they have the same type, and
\item have a similarity greater than a threshold. 
Similarity is computed based on the common number of mapped descendants that both nodes have.
Increasing this threshold implies that a bottom-up matcher increases the minimum ratio of common descendants and, consequently, tries to match more similar subtrees.
\end{inparaenum}
The bottom-up matchers of GumTree obtain the similarity threshold in different ways.
The greedy matcher uses the hyperparameter \texttt{BUM\_SMT} to establish the similarity threshold.
The default value is 0.5, which means that a bottom-up matcher only considers nodes that have, at least, a 50\% of common descendants.
The other two bottom-up matchers (Simple and Hybrid) automatically compute the threshold using the following formula:
$\mathit{threshold}(t_1,t_2) = 1/ (1 + \mathit{log}( desc(t_1) + \mathit{desc}(t_2) )) $, where $t_1, t_2$ are subtrees, and $\mathit{desc}(t)$ gives the number of descendants of subtree $t$. 

\paragraph{Maximum size threshold}

As explained previously, once a bottom-up matcher finds, from a subtree on tree $t_1$,  the most similar subtree from tree $t_2$, 
it applies a \emph{recovery} phase (Section \ref{sec:gumtree:matcher}), which relies on an algorithm that searches for matches among the descendants of both subtrees that are still unmapped.
Classic uses an optimal tree-edit distance algorithm which has a cubic complexity and, therefore, is slow on large subtrees.
Simple uses an heuristic which is much faster than the optimal algorithm.
Hybrid applies the algorithm of classic or simple, depending on the size of the subtree under consideration.
Given the fact that classic and hybrid matchers can have a large running time if they try to apply the optimal tree-distance algorithm on large subtrees, they use the maximum size threshold hyperparameter, which establishes the maximum size of a subtree for which this algorithm is applied. This hyperparameter is called \texttt{BUM\_SZT} and its default value is $1000$.

\subsubsection{Defining the Hyperparameter Domain}
\label{sec:hyperdomain}

Table \ref{tab:hyperparameters}  shows the domain of each hyperparameter space.
Some hyperparameters are numeric; in this case, we give the minimum and maximum values.
In addition, we give a reasonable step value to explore the input domain for this parameter, this value was suggested and agreed on with the GumTree lead developer.
For example, the hyperparameter \texttt{BUM\_SMT} goes from 0.1 to 1, with steps 0.1, giving as a result the hyperparameters \{0.1, 0.2, 0.3, $\dots$, 0.9, 1\}.
For hyperparameters with categorical scale, we give a list of possible values.
For example, the bottom-up matcher hyperparameter could receive three values: \emph{Classic, Simple, or Hybrid}.
The `Total' column gives the number of values to explore per hyperparameter.

We recall that the Cartesian product on all hyperparameters creates all possible \algoconfigs that \autotuningname{} evaluates.
In total, we obtain \totalgtconfigsforjava different configurations in GumTree.
Note that this total is not equivalent to the scalar multiplication of the values of each hyperparameter (shown in the last column of Table \ref{tab:hyperparameters} because there are dependencies between the hyperparameters.
For example, the hyperparameter \texttt{BUM\_SZT} is used by (\emph{GreedyBottomUpMatcher} matcher but not by \emph{SimpleBottomUpMatcher}.
Thus, 2000 configurations correspond to ClassicGumTreeMatcher, 
200 to HybridGumTreeMatcher and 10 to SimpleGumTreeMatcher,

\subsubsection{Metamodels}
In our experiments, we perform AST differencing on Java programs, as done in the original publication of GumTree~\cite{falleri_fine-grained_2014}.
GumTree supports multiple AST meta-models for Java code.
The default one is called \emph{JDT}, it is based on the Eclipse JDT Parser.
The other meta-model we choose is the one defined using Spoon~\cite{pawlak:hal-01078532}, an open-source library to analyze, rewrite, transform, and transpile Java source code.

\section{\autotuningname: An Approach for Hyperparameter Optimization of AST Differencing}
\label{sec:dat}

\label{sec:approach_introduction}

We present \autotuningname, an approach to optimize AST differencing algorithms which have configuration parameters such as \cite{falleri_fine-grained_2014, fluri_change_2007, hashimoto_diffts_2008, dotzler_move-optimized_2016}.
The main goal of \autotuningname is to find the optimal \algoconfigs of an AST differencing algorithm in a data-driven manner. 
\autotuningname finds the best \algoconfigs  with respect to a benchmark of file pairs to diff.

\begin{figure*}[t]
\centerline{
\includegraphics[scale=0.60]{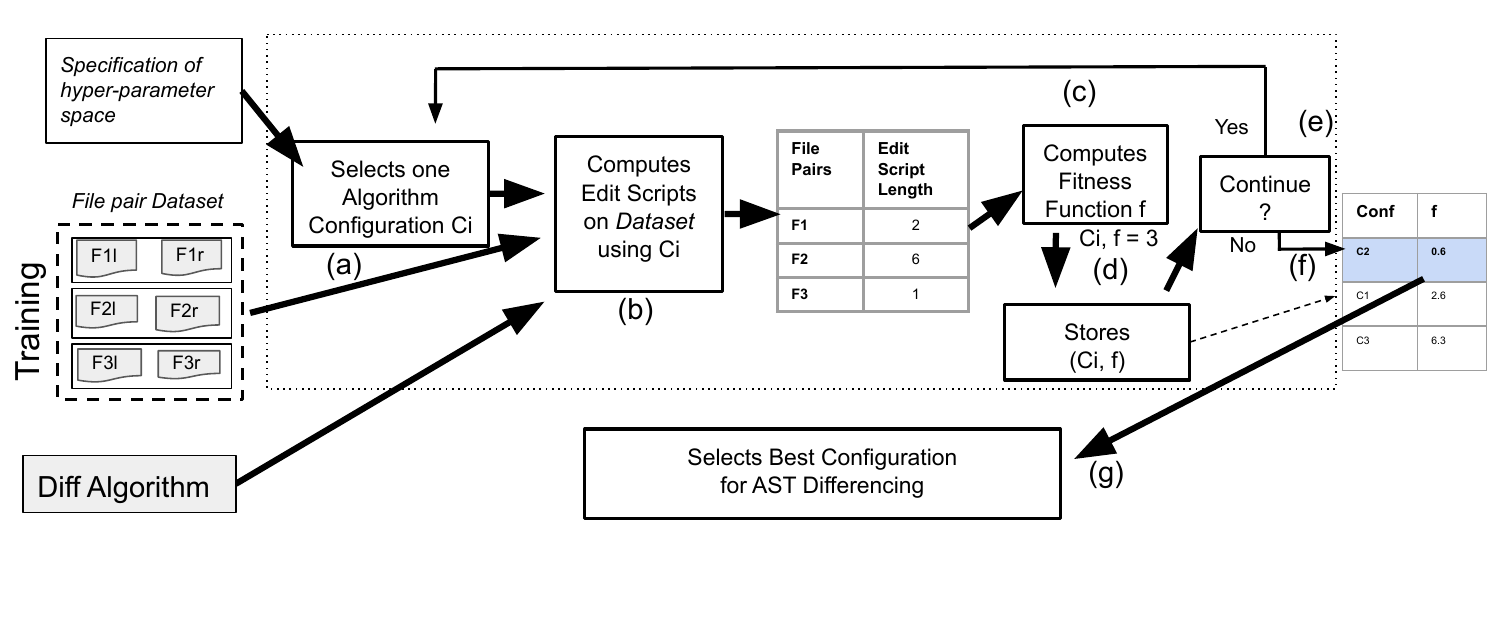}}
\caption{Workflow of \autotuningname{}. It searches for the best AST differencing configuration in a data-driven manner, according to a set of file-pair.}
\label{fig:dat_training}
\end{figure*}

\subsection{AST Differencing Hyperparameters}
\label{sec:diffhyper}

An AST diff hyperparameter is a parameter whose value is used to control a particular procedure from an AST diff algorithm.
A hyperparameter has a domain and a default value.
The \emph{default configuration} of a diff algorithm corresponds to the default values of the set of hyperparameters.
Hyperparameters can be set using different mechanisms:
\begin{inparaenum}[\it 1)]
\item at compile time, by writing the values directly in the code;
\item at startup time, through a command-line option or environment variable;
\item at run time, through a method call.
\end{inparaenum}

To better understand hyperparameters, we take the example of the GumTree algorithm. At some point, GumTree computes a mapping between nodes from the ASTs under comparison that have more than a certain ratio of common children. 
This threshold is called \texttt{BUM\_SMT}, its domain goes from zero to one, and its default value is $0.5$. 
This threshold is an hyperparameter of GumTree.

A \textbf{diff \algoconfig} for an AST diff algorithm $\delta$  is a set of actual values for all available hyperparameters in $\delta$. 
Hyperparameters are noted $h_1, h_2, \dots, h_k$ and hyperparameter values are noted $v_1, v_2, \dots, v_k$.
A value $v_i$ belongs to the \textbf{hyperparameter space} $\text{HS}_i$ for $h_i$. The \textbf{hyperoptimization space} contains the Cartesian product of all hyperparameter spaces.
The hyperoptimization space has $n$-dimensions, where each corresponds to a hyperparameter.
A point in that space corresponds to a particular \algoconfig of a diff algorithm.
For each diff algorithm, there is one single point in its hyperparameter optimization space that corresponds to the \emph{default} configuration.

An important dimension of AST differencing algorithms is the used \emph{AST meta-model}.
Two meta-models can produce, for the same source code file, two different ASTs in terms of topology (such as size or height).
However, thresholds applied to values computed from the topological properties of the ASTs are very common in AST differencing algorithms.
For instance, in GumTree at some point an optional mapping phase is launched depending on the number of children of the node under consideration (the \texttt{BUM\_SZT} hyperparameter which default value is 1000). For instance, as the Spoon Java parser produces ASTs with more nodes than the JDT Java parser\footnote{We applied a statistical test to ensure that ASTs generated from the JDT meta-model (JDT) are different from those generated from the Spoon meta-model.
As the distributions of AST' sizes and heights are not normalized, we use a Wilcoxon signed-rank test to reject (at 0.05 level) the two null-hypotheses which state that the size (resp. height) of a $\mathit{AST}_{\text{Spoon}}$ is similar to the size (resp. height) of an $\mathit{AST}_{\text{JDT}}$.}, it could affect the behavior of the algorithm.
For this reason, we suspect that a first interesting case to apply hyperparameter optimization is at the level of each AST meta-model.

Another important dimension is the source code itself. For instance, a source code with very long methods will also induce a very different behavior of the algorithm, due to the previously mentioned threshold, compared to a source code with very short methods.
Therefore, a second interesting case for hyperparameter optimization is at the level of given file pair for a given AST meta-model.

We summarize these two use cases in the next section.

\subsection{Use-cases of \autotuningname{}}
There are two main use cases of \autotuningname{}:
\emph{global} and \emph{local} hyperparameter optimization.

For \textbf{global hyperparameter optimization}, the goal is to find a default configuration that is expected to work well for ASTs produced using a given meta-model. The global hyperparameter optimization is designed to assist AST differencing tools maintainers to compute good default values for the hyperparameters for a given AST meta-model. 
These values will then be automatically applied when using the diff tool to provide an improved end-user experience without suffering additional cost (the optimized values are computed on a training set and then stored in the diff tool).

For this use case \autotuningname does a \emph{set}-based optimization: it searches for the best-performing configuration over a set of training file pairs written in a same language and parsed using a same meta-model.

For \textbf{local hyperparameter optimization}, \autotuningname does a \emph{case}-based optimization: 
it searches for the best performing configuration on a given file pair, using a given meta-model. It is designed for end-users that want to compute the best possible diff at the expense of computation time (since in this setup the optimization is computed each time an end-user wants to diff a file pair).

\subsection{Workflow}
\label{sec:overview}

\autotuningname takes as input:
\begin{inparaenum}[\it 1)]
\item an \emph{unlabelled} dataset of file pairs,
\item a meta-model,
\item an AST diff algorithm,
\item and the specification of the algorithm's hyperparameter space.
\end{inparaenum}
It outputs a sorted list of \algoconfigs, ordered in ascending order, from the best to the worst performing according to a given metric.

\autotuningname{} implements the workflow presented in Figure~\ref{fig:dat_training}.
Given a set of file pairs (i.e., a training dataset for data-driven optimization), and a specification of the hyperparameter space,
\autotuningname first selects one \algoconfig, $C_i$, to evaluate (step a).
The exploration and selection of other configurations is done according to a search technique (as described in Section~\ref{sec:search_techniques}).

Then, \autotuningname{} computes the edit-scripts for each file pair from the input dataset using the selected configuration $C_i$ (step b).
After having executed all the AST differencing tasks, \autotuningname{} computes the $fitness$ value $C_i$, this fitness value then guides the search process.
As the fitness value, by default \autotuningname{} computes the averages of the length of those edit-scripts (step c), and stores that fitness value together with the configuration in a list (step d).
Note that \autotuningname{} can be extended to use another fitness function.
Then, it may repeat the aforementioned step by selecting another \algoconfig (step e), or it stops the search (step f).
The stopping criterion depends on the search technique employed by \autotuningname{}.

Finally, \autotuningname sorts the list with the evaluated configurations according to their fitness values (i.e., the average edit-script lengths) in increasing order: 
\algoconfigs which produce shorter edit-scripts have better fitness than those \algoconfigs which produce larger edit-scripts on average.

The best \algoconfig found by \autotuningname{} in the previous phase (the first element from the returned list) can be used to test the optimization done by \autotuningname{} on a separate dataset. 
For example, we can compare the performance on a testing dataset of the best configuration found (step g) with that one from the \emph{default} configuration (step h).
This best configuration can be used in production as the new \emph{default} configuration of the AST diff algorithm under optimization.

In Sections \ref{sec:fitnessfunction} and \ref{sec:search_techniques}, we detail, respectively, the fitness function and the search techniques employed by \autotuningname{}.

\subsection{Fitness Function}
\label{sec:fitnessfunction}

The main challenge that AST diff hyperparameter optimization faces is the definition of a good fitness function.
There is no ground truth on whether a hyperparameter configuration is good or not.
\autotuningname overcomes the lack of ground truth for hyperparameter optimization by using metrics that compare the outputs of two \algoconfigs (i.e., two edit-scripts).
Based on the result of that comparison, \autotuningname decides which of those \algoconfig produces an edit-script of \emph{higher quality}.

We use the \emph{length of an edit-script} as the metric to guide hyperparameter optimization. 
As suggested by~\cite{falleri_fine-grained_2014}, the length of the edit-scripts is an indicator of the effort required to understand the changes between two files, because shorter edit-scripts are typically easier to understand.
Moreover, it is an accepted proxy of the quality of the edit-scripts: other researchers have used it to measure the improvements introduced by new AST diff algorithms~(e.g.,~\cite{higo_generating_2017, hashimoto_diffts_2008, huang_cldiff_2018, frick_generating_2018, dotzler_move-optimized_2016, matsumoto_beyond_2019}).

\autotuningname{}'s fitness function takes three parameters: 
\begin{inparaenum}[\it i)]
    \item C: a single algorithm configuration (as defined in Section \ref{sec:diffhyper})
    \item TS: a the training set composed of a set of file pairs. The diff algorithm, configured with C, is executed on each pair from TS.
    \item M: a metric (e.g., average, median) that quantitatively measures the diff quality (i.e., the script length).
\end{inparaenum}

\subsection{Search Techniques}
\label{sec:search_techniques}

\autotuningname{} includes powerful search-based techniques used in hyperparameter optimization for other software engineering tasks (e.g.  \cite{Agrawal2020HyperparameterOptimization}): GridSearch, Hyperopt and Optuna.
It focuses on techniques that are implemented in open-source generic off-the-shelf optimization frameworks.

\subsubsection{Grid Search \cite{Bergstra2011AlgorithmsHyper}}
\label{sec:grdisection}
this optimization method exhaustively searches through a specified slice of the hyperparameter optimization space. 
Given the specification of a hyperparameter space, GridSearch  first creates \algoconfigs by performing the Cartesian product between all the selected subsets of hyperparameters. For each \algoconfigs, it computes the fitness function described in Section \ref{sec:fitnessfunction} on a set of file-pairs (the training set). Finally, it returns the configurations with the highest fitness (the shortest edit script).

\subsubsection{Hyperopt \cite{Bergstra2015Hyperop}} is a distributed asynchronous hyper-parameter optimization framework, based on the Tree-of-Parzen-Estimators (TPE) algorithm \cite{Bergstra2011AlgorithmsHyper}.
TPE is a Bayesian optimization approach~\cite{Shahriari2015}, based on \emph{Sequential Model-Based Optimization}~\cite{Hutter2011smbo}), which builds a probability model $p(score|hyperparameter)$ (aka the ``surrogate'' function) of the objective fitness function and uses it to select the most promising hyperparameters to evaluate in the objective function. 
The advantage of this algorithm is when evaluating expensive fitness functions.
In these cases, w.r.t. Grid Search, it is less expensive because it explores a part of the search space, leading to lower execution costs~\cite{Bergstra2011AlgorithmsHyper}. 
As mentioned by \cite{lustosa2023optimizingSNEAK} Hypeort is the most cited optimization algorithm.

\subsubsection{Optuna \cite{optuna_2019}} is an hyperparameter optimization framework designed by \emph{define-by-run criteria} \cite{optuna_2019} (i.e., it allows the user to dynamically construct the search space\cite{optuna_2019}), and furnished with an efficient sampling algorithm and pruning algorithm.
Optuna is also an extension of TPE.

We do not consider other optimization techniques recently presented, such as Dodge~\cite{Agrawal2019Dodge}, niSneak~\cite{lustosa2023optimizingSNEAK}, NUE\footnote{Nue framework \url{https://github.com/lyonva/Nue}} because, after inspecting their code, we found that they are not reusable and deeply tied to machine learning (ML) models from \emph{scikit-learn}~\cite{scikit-learn}. In this paper, we optimize diffs, not ML models. Thus, the utilization of techniques is not feasible without completely reimplementing them.

\section{Experimental Results}
\label{sec:results}

In this section, 
we first present the data set we use for the evaluation of \autotuningname{}.
Then, we respond to the research questions.

\subsection{Evaluation Dataset}
\label{sec:methodology:dataset}

The evaluation of \autotuningname{} consists in running the hyperparameter optimization GumTree on a set of file-pairs.
We create one as follows.
First, we choose repositories of open-source projects in order to extract revisions of files.
We choose CVSVintage~\cite{cvs-vintage}, a dataset composed of 14 CVS repositories of open-source projects written in Java, because GumTree was initially evaluated on that dataset~\cite{falleri_fine-grained_2014}.  

To create the set of file pairs, we first convert each CVS repository to a GIT repository using the tool \texttt{cvs2git}\footnote{cvs2git: \url{https://www.mcs.anl.gov/~jacob/cvs2svn/cvs2git.html}}.
We were able to successfully convert 13 of the 14 repositories. 
Then, we navigate the history of each GIT repository, commit by commit, and for each one, we store the Java files that have been updated according to the command \texttt{git diff}.
For each file $f$ updated by commit $C$, we create a pair file $(f_p, f)$, where $f_p$ is the previous version of file $f$ (i.e., the version before commit $C$).
All file pairs are available in our appendix.

We study file-pairs that introduce AST changes. 
For detecting them, we compute the edit-script vanilla GumTree on each pair and keep those that have an edit-script longer than zero. 
Due to limited computational resources, we consider a subset of files. For that, we randomly select up to 1000 file-pairs per project. 
Nine projects have less than 1000, so we consider all of them for these projects.
In total, we consider \filepairsexecuted{} file-pairs.

\subsection{RQ1: \rqoptimizationimprovementdefault}

\subsubsection{Protocol for RQ1}
\label{sec:methodology_rqoptimization}

To answer this research question, we execute the \emph{global} hyperparameter optimization from \autotuningname{} on the evaluation dataset described in Section \ref{sec:methodology:dataset}, composed of \filepairsexecuted{} file-pairs. 
We performed this hyper-optimization for the two considered AST meta-models (JDT and Spoon) using the GridSearch technique implemented in \autotuningname.

To minimize the risk of data overfitting, we apply a 10-fold cross-validation.
For each fold, we first generate two sets: training (90\% of the data) and testing (the remaining 10\%). 
Then, we hyperoptimize GumTree using the training dataset with the goal of finding the configuration with the best performance C$_{\text{Best}}$.
We compare the performance of configurations using two metrics: reduction of the
\begin{inparaenum}[\it 1)]
    \item median value (50th percentile) and 
    \item 75th percentile value.
\end{inparaenum}
Then, using the testing dataset, we calculate the performance of: 
\begin{inparaenum}[\it a)]
\item the best configuration (C$_{\text{Best}}$), and
\item the default configuration (C$_{\text{Default}}$).
\end{inparaenum}
Next, we compute the proportion of file pairs where:  
\begin{inparaenum}[\it a)]
\item hyper-optimization (C$_{\text{Best}}$)  improves the edit-script w.r.t. default configuration (C$_{\text{Default}}$) (Metric I),
\item hyper-optimization produces an equivalent edit-script (Metric E),
\item hyper-optimization produces a worse edit-script (Metric W).
\end{inparaenum}
Finally, we report the median of I, E and W over all folds.

We also proceed to a statistical assessment of the results using a Wilcoxon signed rank test against the size of the edit-scripts produced by two different configurations: the default and the best one found using the GridSearch technique.
We use this test since we have no assumption about the distribution of the edit-script sizes.
Our null and alternative hypotheses that we focus on in this RQ are as follows:
\begin{itemize}
    \item $H^1_{\text{null}}$ There is no difference between the median length of the edit-scripts produced using the global and default configuration (alternative $H^1_{\text{alt}}$ the edit-scripts produced using global have a median shorter length than the ones produced using default).
\end{itemize}
We also report the effect size using Rosenthal’s R, whose value varies from 0 (small effect) to close to 1 (large effect).

\subsubsection{Results}
Table \ref{tab:optimizationdefault} presents the results.
It displays three columns that present the percentage of cases from the testing set where hyperoptimized GumTree finds: 
\begin{inparaenum}[\it a)]
\item a shorter edit-script than default GumTree, meaning that the hyperoptimization improves the performance of GumTree (column I),
\item the same length edit-script (column E)
\item a larger edit-script, meaning that the hyperoptimization harms the default configuration of the differencing algorithm (column W).
\end{inparaenum}
The table reports two metrics: 
improvement on the 
\begin{inparaenum}[\it 1)]
    \item median,
    \item \nth{75} percentile  value of edit-script length.
\end{inparaenum}
For a matter of space, we discuss the former metric.

\begin{table}[t!]
\def\arraystretch{1.5}
\caption{(RQ1) 
Comparison between the performance of globally hyperoptimized GumTree and default GumTree.
Columns $Med$ and $75P$ are the \% of improvement of the median and \nth{75} percentile values, respectively.}
\label{tab:optimizationdefault}
\centering
 \begin{tabular}{|c  ||  c:c|r:r|r:r| c|
 } 
 \hline
 Meta- &  \multicolumn{6}{c|}{\% cases where  GridSearch optimization}
 & 
 \\
model &  \multicolumn{2}{c|}{Improves (I)}  & \multicolumn{2}{c|}{Equals (E)} & \multicolumn{2}{c|}{Worse (W) }& p-value
  \\
\cline{2-7}
&Med&\nth{75}P&Med&\nth{75}P&Med&\nth{75}P&\\
\hline
\hline   
JDT  &
21.8&22.5&75.6&72.9&2.5&4.4&$2.2e-16$\\
\hline
 Spoon  &
16.1&16.3&81.8&81.4&2.2&2.1&$2.2e-16$\\
 \hline
 \end{tabular}
\end{table}
 
To diff ASTs from JDT, the optimization of GumTree improves the performance of the default GumTree for {21.8\%} of cases.
The detriment of applying global optimization is much lower: in only 2.5\% of the cases, hyperoptimized GumTree produces larger edit-scripts.
For the rest of the cases (75.6\%), hyperoptimization has no impact on the length of edit-script produced by GumTree.
Table~\ref{tab:Globalexampleconfigs} shows the best configurations found by \autotuningname{} for the JDT and Spoon meta-models, and the default value used by GumTree.
We observe that for JDT the best configuration uses a different matching algorithm (Hybrid) than the default configuration (Classic).

We use a Wilcoxon signed rank test to statistically assess the differences in edit-script lengths produced using the hyperoptimized configuration versus the default configuration.
The obtained P-value is $2.2e-16$, therefore, we reject the null hypothesis $H^1_{null}$. The effect size, calculated using Rosenthal's R, is$-0.585$, which can be considered between medium and large.

To diff ASTs designed with the Spoon meta-model, hyperoptimization with \autotuningname has less impact on the number of improved cases (16.1\%).
It means that the default configuration of GumTree works already well in most cases.
We observe from Table~\ref{tab:Globalexampleconfigs} that the best configuration for Spoon has the same matching algorithm as the default (Classic). 
However, there are three parameters that receive different values: STM\_ PC, BUM\_SMT, and BUM\_SZT.

Again, we run a Wilcoxon signed rank test on the edit-script length distribution and the obtained P-value is inferior to $2.2e-16$, therefore, we reject the null hypothesis $H^1_{null}$. The effect size, computed using Rosenthal's R {is $-0.669$, which can be considered between medium and large.}

\begin{figure}[t]
\centering
\includegraphics[width=0.9\columnwidth]{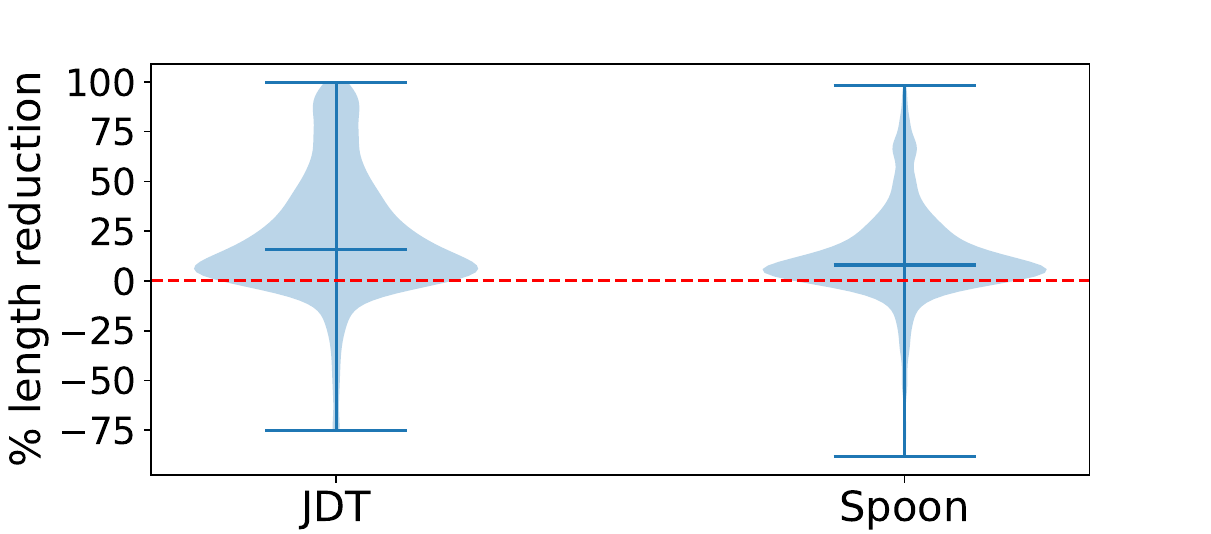}
 \caption{Distribution of the percentage of reduction of the edit-script length using the optimized configuration w.r.t. the default configuration. (Cases with no improvement or detriment are ignored).
 }
 \label{fig:distributionPergentage}
\end{figure}

\begin{figure}[t]
\centering
\includegraphics[width=\columnwidth]{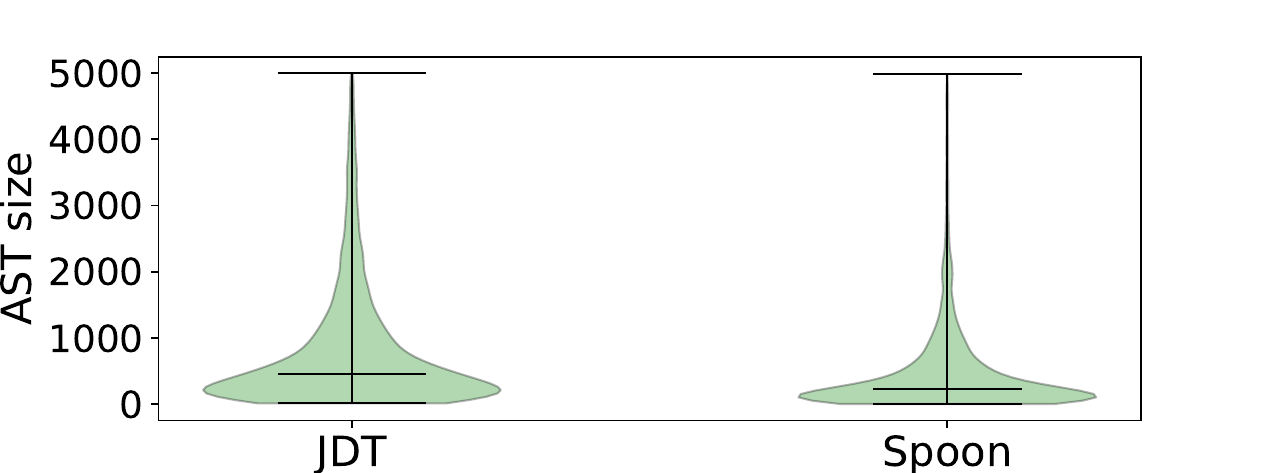}
 \caption{Distribution of AST sizes (\# of nodes).}
 \label{fig:distributionSizes}
\end{figure}

Figure~\ref{fig:distributionPergentage} shows the distribution of the percentage of reduction in the size of the edit-scripts calculated with the optimized configuration compared to the edit-script calculated with the default configuration. 
The figure considers the cases that present improvement due to the optimization process (these have \% positive and correspond to 21.8\% in JDT and 16.1\% in Spoon), cases for which optimization produces worse results (\% negative, 2.5\% in JDT and 2.2\% in Spoon), but ignores those with equal results to facilitate visualization.
For JDT, half of the optimization-affected cases have a reduction of at least 20\%, and there is a considerable number of cases with an improvement greater than 75\%.
In contrast, for Spoon, the reduction in the size of the edit-scripts is less than for JDT, and there are fewer cases that are reduced by more than 50\%.
That can be explained by the fact that Spoon ASTs are smaller. 
Figure \ref{fig:distributionSizes} shows the distribution of the AST sizes (expressed in \#~of nodes).
The JDT ASTs have a median of 480 nodes, while Spoon ASTs have 235.
Consequently, a change on Spoon ASTs typically spans a smaller number of nodes than on JDT.

We also inspected cases in which the tuning resulted in longer edit scripts.
We observe that this happens when a file-pair under analysis has an insertion of code that already exists in the file. 
The tuned version of GumTree creates spurious moves and additional inserts related to this code.

In conclusion, this experiment shows that the effect of the optimization is different according to the meta-model used: for JDT, the optimization 
\begin{inparaenum}[\it a)]
\item affected more cases, and 
\item for those affected cases, the reduction is more significant.
\end{inparaenum}

\begin{tcolorbox}[boxsep=0.0pt]
{\bf \underline{Answer to RQ1:}}
Global hyperparameter optimization improves GumTree performance by providing a better configuration than the default configuration. The hyperoptimized configuration produces the shortest edit-scripts for 21.8\% of the cases using the JDT meta-model and for 16.1\% of the cases based on using the Spoon meta-model, in a statistically significant manner. 
\end{tcolorbox}

\begin{tcolorbox}[boxsep=0.0pt]
{\bf \underline{Implications for practitioners:}}
Maintainers and users of AST differencing tools such as GumTree have a new tool in their toolbox. When they apply AST differencing to a new programming language, they can first perform a global hyperparameter optimization, which would identify a new configuration that is 
\begin{inparaenum}[\it 1)]
\item  better than the default,  and 
\item tuned to a given AST meta-model.
\end{inparaenum}
\end{tcolorbox}

\begin{table}[t]
\centering

\def\arraystretch{1.2}
\begin{tabular}{|l|c|c|c|}
\hline
\textbf{Hyperparameter} & \textbf{Default} & \textbf{Best for JDT}  & \textbf{Best for Spoon} \\ 
\hline
\hline
Matching Algorithm& Classic& Hybrid &Classic\\
\hline
STM\_ PC& Height& Size &Size\\
\hline
STM MPTH&  1& 1&1\\
\hline
BUM\_SMT&0.5& - &0.2\\
\hline
BUM\_SZT&1000& {400} &600\\
\hline
\end{tabular}
\caption{(RQ 1) Global configurations found by \autotuningname{} using GridSearch for JDT and Spoon Java meta-models. Symbol `-' means that the configuration does not use the hyper-parameter.
}
\label{tab:Globalexampleconfigs}
\end{table}

\subsection{RQ2:  \rqtechnique}
\label{sec:evalrqcost}

\subsubsection{Protocol for RQ2}
\label{sec:protocoltechnique}

In this research question, we evaluate two other hyper-parameter optimization techniques provided by \autotuningname{}: Hyperopt~\cite{Bergstra2015Hyperop} and Optuna~\cite{optuna_2019}.
Rather than evaluating all \algoconfigs as GridSearch does (in total, 2210 as described in Section \ref{sec:hyperdomain}), these techniques are able to perform fewer evaluations. This number is given to Hyperopt and Optuna as input parameters. 
In this experiment, we set up 100 evaluations, which corresponds to the default value from Hyperopt.
Doing 100 evals per diff means that Hyperopt and Optuna will do just the 4.5\% of the evaluations that GridSearch performs per diff.

\subsubsection{Results}

\begin{table}[t]
\def\arraystretch{1.5}

  \centering
  \begin{tabular}{|c| r| |c|c|| c|c|}
    \hline
\multirow{2}{*}{Method }  &  {Evals}  & \multicolumn{2}{|c|}{JDT}  &     \multicolumn{2}{|c|}{Spoon}   \\ 
\cline{3-6}
    &/diffs& Med &  \nth{75}P & Med &\nth{75}P  \\
    \hline
    GridSearch & 2210 & 21.9  & 22.5   & 15.9  &16.24    \\ \hline
    Hyperopt & 100 & 20.2  & 22.8  & 15.2  & 16.13 \\ \hline
    Optuna & 100 & 20.2 & 22.8 & 15.8  & 16.07\\ \hline
    
  \end{tabular}

\caption{
(RQ2) Percentage of improvement of GumTree using the best configurations found by \autotuningname{} GridSearch, Hyperopt, and Optuna. It shows the improvement of the median value (column $\mathit{Med}$) and on the 75th percentile ($\nth{75}P$)}.
\label{tab:globaltpe}

\end{table} Table \ref{tab:globaltpe} shows the percentage of improved cases of GumTree using the best configuration found by GridSearch, Hyperopts and Optima compared to the default configuration of GumTree. We present the results from diffs computed on JDT and Spoon ASTs.
The column $\mathit{Eval}/\mathit{diffs}$ shows the number of diffs executed by each technique in each file-pair from the training dataset. We recall that each execution has a different \algoconfig. 
The column $\mathit{Med}$ means the percentage of cases from the testing set where the best configuration reduces the median value of the fitness function (in our case, the length of the edit-script), and the column $\mathit{75thP}$ the percentage of cases where the best configuration reduces the 75th percentile value of the fitness function.

First of all, we observe that the values of the two metrics using any of the optimization techniques are similar, for the two meta-models.
For JDT, the median value obtained using GridSearch is slightly higher than the value obtained using Hyperopt and Optima: 21.9\% and 20.2\%. However, the number of executions per diff that the latter requires is much lower: 100 vs. 2210.
This means that, by using Hypeorpt or Optuna, one can optimize GumTree with much less computational effort.
For JDT, we do not observe any significant differences between Hyperopt and Optuna on the two metrics.

For Spoon, we observe that:
\begin{inparaenum}[\it a)]
    \item the percentage of improvement is lower than that one obtained for JDT ($\approx$20\% vs $\approx$15\%),
    \item  Optuna has almost the same rate of improvement as GridSearch for the median metric (15.9\% vs 15.8\%),
    \item 
    Optuna reports a higher median value (15.8\% vs 15.2\%), but Hyperopt reports a higher improvement on the \nth{75} percentile (16.13\% vs 16.07\%).
\end{inparaenum}

\begin{tcolorbox}[boxsep=0.0pt]
{\bf \underline{Answer to RQ2:}}
Using Hyperopt and Optuna, \autotuningname can significantly reduce the number of evaluations performed to find the best configuration of GumTree, resulting in faster hyperoptimization.
\end{tcolorbox}

\begin{tcolorbox}[boxsep=0.0pt]
{\bf \underline{Implications for practitioners:}} To perform a global hyperparameter optimization for a new programming language or meta-model, it is enough to collect less than 10000 file-pairs (diffs) and use Hyperopt or Optuna.
This optimization is done only once for a given programming language and meta-model.
\end{tcolorbox}

\subsection{RQ3: \rqlocaloptimization}
\label{sec:evallocal}

\subsubsection{Protocol for RQ3}

To answer this research question, we extend the protocol used to answer RQ1 (Section \ref{sec:methodology_rqoptimization}) to find the best local configuration for each file pair from the training set. As before, we use GridSearch, Hyperopt and Optuna as optimization techniques.
In particular, for each of those file pair $P=(f_p, f)$, we compare the fitness value (i.e., length of the edit-script) given by the local search on $P$ with the value obtained using the default configuration on $P$.

We also proceed to a statistical assessment of the results using a Wilcoxon signed rank test against the size of the edit-scripts produced by the different configurations.
Our null and alternative hypotheses that are studied in this research question are:
\begin{itemize}
    \item $H^2_{\text{null}}$: There is no difference between the median length of the edit-scripts produced using the local and the default configuration (alternative $H^2_{\text{alt}}$ the edit-scripts produced using length have a median shorter length than those produced using the default).
\end{itemize}

\subsubsection{Results}
\begin{table}[t!]
\centering
\def\arraystretch{1.5}
\begin{tabular}{|l || r | r| r | r|r |} 
\hline 
\multirow{2}{*}{Comparison}  &  \multicolumn{2}{c|}{\% Improved (I) }  &  \multicolumn{2}{c|}{\%  Equal (E) }  
 \\
\cline{2-5}
&JDT&Spoon &JDT&Spoon
\\ 
\hline 

\hline
{\autotuningname{}  Local GridSearch} vs Default& {27.4} & {18.7} &  {72.5} & {81.3} 
\\ 
\hline
{\autotuningname{} Local Hyperopt} vs Default& {26.1} & {16.7}  & {73.8}  & 
{83.3} 
\\

\hline
\autotuningname{} Local Optuna vs Default& {26.1} & {17.3}  & {73.8}  & {82.6} 
\\

\hline

\end{tabular}

\caption{(RQ3) Percentages of cases improved by applying local optimization with \autotuningname{}
vs. default configuration.}
\label{tab:localglobal}
\end{table} 
\begin{table}[t]
\centering

\def\arraystretch{1.2}
\begin{tabular}{|l||r|r|r|}
\hline
\textbf{Hyperparameter} & \textbf{Default}  & \textbf{JDT} &  \textbf{Spoon}\\ 

\hline
Matching Algorithm& Classic&{Classic} &Classic\\
\hline
STM PC& Height& Size &Size\\
\hline
STM MPTH&  1& 1&1\\
\hline
BUM\_SMT&0.5 & 0.1&0.1\\
\hline
BUM\_SZT&1000 & 1100&1000\\
\hline
\end{tabular}
\caption{(RQ 3) Most frequent local hyperoptimized configurations found by \autotuningname{}  for the JDT {and Spoon} meta-models.
}

\label{tab:exampleconfigs}
\end{table} 
Table~\ref{tab:localglobal} presents the results of this experiment.
It contains two sets of columns that present, for both the JDT and Spoon meta-models, the percentages of file-pairs for which hyperoptimization improves the results (I) or produces equal results (E). 
We remark that local optimization from \autotuningname{} never worsens the performance of AST differencing. This is by construction, as it optimizes a single file-pair, if the technique does not find a configuration that improves the default, then the default is used.

The row {\bf\autotuningname{}  Local GridSearch vs. Default} presents the results obtained from local optimization.
Local optimization positively impacts the performance of GumTree, producing shorter edit-scripts than those from the default configuration for 27.4\% of cases with JDT and for 18.7\% of cases with Spoon. 
We note that the improvement in local hyperoptimization is larger than that provided by global hyperoptimization ($27.4\%  >> 21.8\%$). 
Using the Wilcoxon signed rank test, we reject the null hypothesis $H^2_{\text{null}}$ for both the JDT and the SPOON meta-models. 
The effect sizes computed using Rosenthal's R are -0.579 and -0.668  respectively, which can be considered between medium and large.

The last two rows in Table~\ref{tab:localglobal} present the results of local optimization using Hyperopt and Optuna.
First, we observe that using these techniques, \autotuningname{} is still able to locally improve the edit-script, but the percentage of improved cases is a bit lower than those using GridSearch achieve almost the same improvement that using GridSearch (e.g., for JDT, Hyperopt achieves  26.1\% vs 27.4\%).
However, the number of evaluations (i.e., executions of diff) per file-pair under comparison is much lower using Hyperopt or Optuna: 100 for Hyperopt-TPE (default number of evaluations) vs. 2210 for GridSearch (we recall that it does an exhaustive search on the configuration space).
Second, we observe that Optuna achieves slightly better results than Hyperopt on Spoon, and similar on JDT.
We have observed this trend in global optimization (Section \ref{sec:evalrqcost}).

Developers and practitioners can decide the search method used by \autotuningname{} according to the scenario they apply \autotuningname{}, for example, a scenario with a limited budget or where they need to optimize fast, thus techniques such as Hyperopt or Optuna, both based on TPE, would be more convenient, or another where they need to obtain the best result and do not have budget restrictions, thus GridSearch would be a better option.

\begin{table*}[th]
\centering
\def\arraystretch{1.2}
\begin{tabular}{|c|r|r|r|r|r||r|r|r|r|r|}
\hline
\#                     & \multicolumn{5}{c||}{JDT}    & \multicolumn{5}{c|}{Spoon}  \\ \cline{2-11}
\multirow{2}{*}{evals}                  & \multicolumn{1}{c|}{Diff}   & \multicolumn{2}{c|}{Hyperopt}    & \multicolumn{2}{c||}{Opt}                                  & \multicolumn{1}{c|}{Diff}   & \multicolumn{2}{c|}{Hyperopt}                                                  & \multicolumn{2}{c|}{Opt}            \\
\cline{2-11}
\multicolumn{1}{|c|}{} & \multicolumn{1}{c|}{t}      & \multicolumn{1}{c|}{t}     & \multicolumn{1}{c|}{I}   & \multicolumn{1}{c|}{t}     & \multicolumn{1}{c||}{I}   & \multicolumn{1}{c|}{t}      & \multicolumn{1}{c|}{t}     & \multicolumn{1}{c|}{I}   & \multicolumn{1}{c|}{t}     & \multicolumn{1}{c|}{I}   \\
\hline
100                    &  \multirow{4}{*}{0.065s} & {9.85s} & {27}  & {8.71s} & {27} &  \multirow{4}{*}{0.027s} & {4.01s} & {15}  & {3.58s} & {15}\\
\cline{3-6}
\cline{8-11}
50                     & {}       & {6.5s}  & {27}  & {5.94s} & {27} & {}       & {2.97s} & {15}  & {2.8s}  & {14}  \\
\cline{3-6}
\cline{8-11}
25                     & {}       & {2.8s}  & {27}  & {2.24s} & {27}  & {}       & {1.54s} & {12}  & {1.19s} & {14}  \\
\cline{3-6}
\cline{8-11}

10                     & {}       & {1.39s} & {26} & {1.02s} & {23} & {}       & {1.05s} & {12}  & {0.89s} & {9}   \\
\hline
\end{tabular}
\caption{{(RQ 4) Execution times of \autotuningname{} using Hyperopt and Optuna in seconds (columns \texttt{t}). The columns `Diff' show the median time for executing a single diff. Columns  \texttt{I} show the percentage of cases improved using local optimization w.r.t. the default configuration.}}
\label{tab:executionLocal}
\end{table*} 
\begin{tcolorbox}[boxsep=0.0pt]
{\bf \underline{Answer to RQ3:}}
Local hyperparameter optimization for a single differencing task is doable: it allows practitioners to find shorter edit-scripts than the default configuration 26.1-27.4\% of cases for the JDT meta-model and 16.7-18.7\% for the Spoon meta-model.
\end{tcolorbox}

\begin{tcolorbox}[boxsep=0.0pt]
{\bf \underline{Implications for practitioners:}}
In sensitive downstream tasks, 
we strongly advise using local hyperparameter optimization for AST differencing in order to obtain the best edit-script according to a fitness function specific to the downstream task.

\end{tcolorbox}

\subsection{RQ4: \rqexecutiontime}

\subsubsection{Protocol}
We focus on HyperOpt and Optuna because, as shown previously in RQ1, they achieve similar performance compared to GridSearch but execute fewer evaluations.
We start running them using the default number of evaluations  (100 evals).
We compute, for each file-pair, the execution time of hyper-optimizing and the execution time of computing a diff using vanilla GumTree (i.e., using the default hyper-parameters) and then report the median values.
Then, we measure the impact of reducing the number of evaluations on execution time and \% of improvements.
We run both HyperOpt and Optuna using 10, 25 and 50 evals per file-pair.
Due to the large magnitude of our experiment (2 tools X 4 \# of evals), we execute the whole experiment on 100 file-pairs randomly selected from our dataset.

\subsubsection{Results}

Table \ref{tab:executionLocal} shows the median execution times (in seconds) of:
\begin{inparaenum}[\it a)]
    \item executing one diff,
    \item execution local optimization using \autotuningname{} Hyperopt and Optuna.
\end{inparaenum}

We first focus on row ``\#evals 100", which is the default value.
We observe that the median diff time takes a few milliseconds (0.065 and 0.027 seconds for JDT and Spoon, respectively).
Running hyper-optimization using 100 evals requires less than 10 seconds.
This includes the time for: 
\begin{inparaenum}[\it 1)]
    \item determining the configurations to execute on each step, and 
    \item executing 100 diffs using these configurations.
\end{inparaenum}
We also observe that hyper-optimizing the Spoon metamodel is faster than optimizing JDT. 
The reason is that executing a diff on a Spoon AST is faster than on its corresponding JDT AST. This happens because, as shown in Figure~\ref{fig:distributionSizes}, Spoon ASTs are smaller than JDT ASTs.

As this optimization time could be too long in practice, we try lower numbers of evaluations per diff: 10, 25 and 50.  
Table \ref{tab:executionLocal} shows the median time to optimize the diff using these numbers of evaluations (columns \texttt{t}) and the number of diffs improved by local optimization (columns \texttt{I}).
We find that using less evaluations, in particular 25, both Hyperopt and Optuna:
\begin{inparaenum}[\it a)]
    \item keep the acceptable performance w.r.t 100 evaluations (in JDT, \autotuningname{} has the same \% of improvement, while for Spoon this percentage is slightly reduced), 
and 
\item reduce the optimization time of a diff to less than 2 and 3 seconds for Spoon and JDT, respectively.
\end{inparaenum}

Furthermore, using 10 evaluations, local optimization is capable of working, in a median, below 1.5 seconds for HyperOpt and below $\approx$1 second for Optuna and is still capable of producing improvements on diffs. 
The reduction in the number of evaluations impacts more on Optuna than on Hyperopt.
For Hyperopt in JDT, there is only one case (out of the 27 improved using 25 evals) that is not improved.
In contrast, Optuna cannot improve 4 cases (out of the 27 improved using 25 evals) on JDT and 5 cases (out of the 14 improved using 25 evals).

These setups (10 and 25 evals) make \autotuningname{} usable in practice, in our opinion.
As the number of evaluations is a hyperparameter of \autotuningname{} local optimization, the user can adjust it according to his/her requirements: either they want much better diffs or fast results.

\begin{tcolorbox}[boxsep=0.0pt]
{\bf \underline{Answer to RQ4:}}
Local hyperparameter optimization, using Optuna with 25 evaluations, takes a median overhead of 1.19 and 2.24 seconds per diff for Spoon and JDT meta-models, respectively.
Moreover, \autotuningname{} using Optuna for local optimization is faster than using Hyperopt.
\end{tcolorbox}

\begin{tcolorbox}[boxsep=0.0pt]
{\bf \underline{Implications for practitioners:}}
In tasks such as visual representations of an AST diff (e.g., using the GumTree CLI), local optimization provides developers with an optimized AST diff within a few seconds.

\end{tcolorbox}

\subsection{Discussion about trade-off between local and global optimization}
\label{sec:discussion:tradeoff}

We consider that global and local optimization can be used in a complementary way and none replaces the other.
Global optimization is designed to be executed when a new language or metamodel is introduced to an AST differencing algorithm, with the goal of learning the new default configuration for that novel language.
Also, global optimization can be used when an AST diff algorithm is deployed in the infrastructure of an organization. There, global optimization is executed on existing code (e.g., code repositories) from that organization in order to learn its peculiarities. 
For example,  limiting the length of methods in order to avoid the code smell `long method' will impact on the size and length of the ASTs from that code. Hyperoptimizing the parameters that consider these variables (e.g. \texttt{BUM\_SZT}) would potentially lead to an improvement in the AST diffs.

In contrast,
we consider that an appropriate use case for local optimization is when a developer detects anomalies on a diff (e.g., a corner case, potentially generated using the global configuration) and wants to improve it.
There, the local optimization is executed \emph{on demand} and it is fast enough.
To that extent,  local optimization is not meant to replace global optimization but to complement it.

\subsection{Analysis of the cases studies}

In this section, we discuss how the best global hyperparameter configuration found by \autotuningname helps GumTree produce a different output from the default configuration in each of the cases presented in Section~\ref{sec:motivation}.

\subsubsection{Case 1: Spurious Add-Remove}
\label{sec:results:case1}

As described in Section \ref{sec:motivation:case1}, GumTree, using both JDT and Spoon meta-models, produces six spurious edits that must not be produced.
GumTree using the best configuration found by \autotuningname does not produce them.
The reason for having different edit-scripts when GumTree uses default and the best configuration is the following.

The three AST nodes in Section \ref{sec:motivation:case1} cannot be mapped during the first phase (top-down matching explained in Section \ref{sec:gumtree:matcher}). 
Both Classic (default) and Hybrid (the matcher used by the best configuration found by \autotuningname) matchers apply the same top-down matching strategy: Greedy subtree matching.
In the second matching phase (bottom-up), the Hybrid matcher is able to map those nodes during \emph{recovery phase} (a last step done by a matcher which tries to map the unmapped children of two subtrees whose roots are mapped). 
However, the Classic matcher does not map them because it invokes \emph{recovery phase} only if the size of the trees to match is smaller than the hyperparameter \texttt{BUM\_SZT} (by default 1000).
The size of the parent tree of these three mentioned nodes (which corresponds to the class \texttt{Interpreter}) is greater than \texttt{BUM\_SZT}=1000, so the \emph{recovery phase} is never called and the three nodes remain unmapped.

\subsubsection{Case 2: Including updates in the edit-script}
\label{sec:results:update}

\begin{figure}[t]
\centering
\caption{(Case 2) Visualization of edits by GumTree (two updates) between the file \texttt{PanelWindowContainer.java} (jEdit project) from commit 6867bd (right) and its previous version (left) using  the best global configuration.}
\includegraphics[width=1\columnwidth]{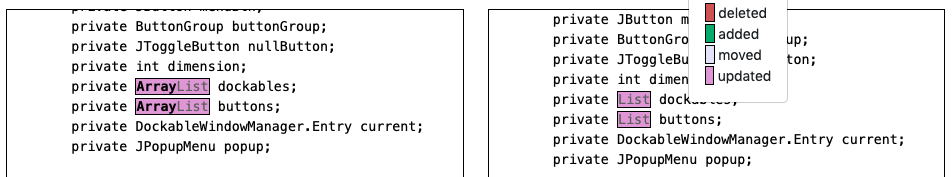}
\label{fig:case:update:best}
\end{figure}

The edit-script generated by GumTree using the best configuration found in Section~\ref{sec:results} is much shorter and understandable: as Figure~\ref{fig:case:update:best} shows, it only includes one update of each field declaration statement rather than adding and removing operating.

During the top-down matching phase using the default configuration, GumTree maps two nodes (one from the left tree, the other from the right tree) if their labels are equal and, following the default values of hyperparameters \texttt{STM\_PC} and \texttt{STM\_MPTH}, their heights are $>=$ 1. 
Note that using this configuration, all leaf nodes (even those that are not modified) from the left tree are not mapped to any from the right tree because their heights are 0. 
Then, during the bottom-up matching match, GumTree cannot match any of the nodes corresponding to the field declaration statements \texttt{private ArrayList dockables;} and \texttt{private ArrayList buttons;} from the left part with those corresponding to the field declarations from the right (\texttt{private List dockables;} and \texttt{private List buttons;}).
This is because the similarity value between the field declarations from the left and right (e.g., \texttt{private ArrayList dockables;} and \texttt{private List dockables;}, respectively) is 0.4, which is lower than the default threshold \texttt{BUM\_SMT} (i.e. 0.5), which controls the mapping of two nodes. 
This low similarity value is due to the fact that most of the descendants have not been mapped. 
As a result, the edit-script generated using the algorithm of Chawathe et al.~\cite{Chawathe1996} (Figure~\ref{fig:case:update:default}) includes remove and add edits that affect these unmapped nodes, including the field declaration.

GumTree tuned with the best configuration, which uses the Hybrid Bottom-up matcher, finds the expected edit-script (Figure \ref{fig:case:update:best}). 
When the tuned version of GumTree uses `size' $=1$, it arrives to match leaves nodes that are not mapped using the default configuration (`height' $=1$).
The map is possible because the size of a leaf node is 1.
These mappings produced by GumTree with the tuned version of GumTree but not with the default impact bottom-up matching: the similarity score between the declaration statements (e.g., \texttt{private ArrayList dockables;} and \texttt{private List dockables;}) is 0.6, higher than the threshold \texttt{BUM\_SMT} with value 0.5. 
For this reason, both statements are mapped, and the edit-script generator does not generate spurious add and remove edits.

\section{Threats to validity}
\label{sec:threatstovalidity}

\subsection{Construct validity}

\emph{Fitness function:} The fitness function is meant to measure the quality of an edit-script.
We decided to use the length of the editing scripts as a fitness function because it has previously been used in the literature for this purpose~\cite{falleri_fine-grained_2014, higo_generating_2017, hashimoto_diffts_2008, huang_cldiff_2018, frick_generating_2018, dotzler_move-optimized_2016, matsumoto_beyond_2019}. Other fitness functions might yield different outcomes.

\subsection{Internal Validity}

\emph{Hyperoptimization techniques:}

\autotuningname{} provides optimization using GridSearch, Hyperopts, and Optuna, which have been proven to be successful in hyperparameter optimization for software engineering tasks \cite{Li2020CPDP,Tantithamthavorn2019ImpactOptimization, lustosa2023optimizingSNEAK}.
There may exist other search techniques that produce better results.
Recall that GridSearch, Hyperopts, and Optuna
are the only generic, off-the-shelf implementations publicly available.
Others, including Dodge~\cite{Agrawal2019Dodge} or niSNEAK~\cite{lustosa2023optimizingSNEAK}, provide implementation tied to machine learning frameworks such as scikit-learn, which makes them unusable for optimizing AST differencing.

\emph{Quantization of the hyperparameter space:} We quantize the hyperparameter space of GumTree using \emph{initial}, \emph{end} and \emph{step} values (all presented Table \ref{tab:hyperparameters}). It could be the case that there are values not included in the selected subset that produce better values.
As Bergstra~\cite{Bergstra2012RandonHyper} remarks, zones of useful hyperparameters might be very narrow within the full search space. Consequently, considering large steps may not help \autotuningname{} to find such zones.
To test whether that happens, we replicate our experiment from RQ4, this time running Hyperopt using a more refined hyperparameter space:
we change the steps of hyperparameter BUM\_SMT from 0.1 to 0.01 and of BUM\_SZT from 100 to 10.
Using Hyperopt with 25 and 100 evals per diff, we observe that the refinement does not improve the results: It either 
\begin{inparaenum}[\it a)]
    \item maintains the same performance, or
    \item reduces the performance up to 2\%. This happens because the refinement enlarged the search space.
\end{inparaenum}

\subsection{External Validity}

\emph{AST metamodeling:}
Our results show the importance of AST meta-modeling in differencing. We selected two meta-models for modeling Java ASTs. We are aware that there exist other meta-models (e.g., JavaParser).
As shown by the clear performance difference between JDT and Spoon, it may happen that hyperparameter optimization by \autotuningname{} performs differently for other Java meta-models.

\emph{Selection of Java:} Our results are made on AST differencing for Java programs, as in most related work on AST differencing analysis \cite{falleri_fine-grained_2014,fluri_change_2007,frick_generating_2018, huang_cldiff_2018}. Future experiments will improve the external validity in other programming languages.

\emph{Evaluation dataset}: We evaluate \autotuningname{} on a dataset composed of 14 popular open-source projects, because GumTree, the algorithm that we hyper-optimize, was initially evaluated on the same dataset.
Future experiments will improve the external validity in other datasets.

\section{Related Work}
\label{sec:relatedwork}

\subsection{Advanced AST Differencing Algorithms}

Since the preliminary work by Chawathe et al. \cite{10.1145/233269.233366}, two AST differencing algorithms with a large impact recently are GumTree~\cite{falleri_fine-grained_2014} and ChangeDistiller~\cite{fluri_change_2007}.
Several works have extended ChangeDistiller and GumTree with the goal of improving their performance.
For example, Higo et al.~\cite{higo_generating_2017} extend GumTree by adding a new edit type: copy-and-paste.
They found that 18\% of the edit-scripts generated using their approach are shorter than those from GumTree.
Dotzler et al.~\cite{dotzler_move-optimized_2016} present MTDIFF, based on ChangeDistiller, which improves the detection of moved code and is able to reduce the length of edit-scripts.
Matsumoto et al. \cite{matsumoto_beyond_2019} present a hybrid AST diff approach that matches the AST nodes with information from the \emph{diff} command (based on the Myers algorithm \cite{myers1986ano}), and is able to generate shorter edit-scripts than \gumtree for the 30-50\% of the cases analyzed.
Yang and Whitehead \cite{Yang2019Pruning} use textual-differencing to prune the AST. This reduces both the number of nodes and the execution time. 

Frick et al. \cite{frick_generating_2018} present an extension of GumTree, Iterative Java Matcher (IJM), which produces more accurate and compact edit-scripts by optimizing move and update edits.
Huang et al. present ClDiff \cite{huang_cldiff_2018}, an AST differencing algorithm that produces concise edit-scripts by grouping and linking AST nodes affected by changes. It generates shorter edit-scripts  than GumTree for 48\% file-pairs analyzed.

De la Torre et al.~\cite{delaTorre2018Imprecision} study the quality edit-script generated by GumTree and propose four categories of imprecisions on edit-scripts.
They empirically study the presence of such imprecision in C\# programs.
Fan et al.~\cite{fan2021differential} define an approach that calculates statements with inaccurate mappings for AST differencing algorithms. 
They found that GumTree generates inaccurate mappings for 20\%-29\% of the cases.

We have the same goal as most of these papers: reducing the length of the edit-scripts. Yet, even if some of this related work does manual tweaking of hyperparameters, none of those papers does automated hyperoptimization.
Our experiment in this paper shows that the default configuration is indeed suboptimal.
We note that our contribution to AST differencing hyperoptimization is applicable to all the past and future AST differencing algorithms to come.

\subsection{Hyperparameter Optimization in Machine Learning and Software Engineering}

Hyperparameter optimization has been applied in various areas, notably in machine learning.
For example,  Kotthoff et al.  \cite{Kotthoff2016Autoweka} presented Auto-Weka, a tool that automatically applies hyperparameter optimization to machine learning models from Weka. Feurer et al.~\cite{Feurer2015AutoML} applies hyperparameter optimization to models of the Sklearn framework.

Now, let us focus on hyperparameter optimization in software engineering. 
Tantithamthavorn et al. \cite{Tantithamthavorn2019ImpactOptimization}
studied the impact of automated parameter optimization on defect prediction models.
Their study shows that automated parameter optimization can have a large impact on performance stability and interpretation of defect prediction models.
Similarly, Li et al. \cite{Li2020CPDP} studied show that parameter optimization improves the performance of cross-project defect prediction techniques.

Arcuri and Fraser \cite{Arcuri2013ParametertuningSE} applied hyperparameter optimization on EvoSuite \cite{Fraser2011Evosuite} a framework for test generation. 
They showed that this positively impacts the performance of EvoSuite. 
However, the author, as also found by Kotelyanskii et Kapfhammer~\cite{Kotelyanskii2014}, stressed that the parameter settings obtained may be worse than arbitrary default values. 
This has motivated us to precisely measure those cases where hyperoptimization produces worse results than the default configuration of GumTree.
Also related to software testing, Jia et al. \cite{Jia2015Learing} used hyperparameter optimization for Combinatorial Interaction Testing (CIT).

The width of the software engineering domains where hyperoptimization has been used is large.
Apel et al. \cite{Apel2012MegeAutotuning} integrated hyperparameter optimization in JDime, a tool for structured merge. 
Shu et al.~\cite{Shu2021SecurityReports} applied hyperparameter optimization to improve data preprocessing for software bug report classification.
Basios et al.  \cite{Basios2018Darwinian} applied optimization to the problem of selecting data structures that share a common interface.
Panichella \cite{PANICHELLA20211LDAHyper} carried out hyper-optimization of LDA  (Latent Dirichlet Allocation) applied to identify duplicate bug reports.
\cite{Villalobos2021} applied 12 hyperparameter optimization techniques to the software effort estimation task.

Other works have presented hyperparameter optimization techniques that were evaluated in software engineering tasks.
Agrawal et al.~\cite{Agrawal2018SMOTUNED} presented SMOTUNED, an approach that tunes SMOTE~\cite{Chawla2002SMOTE}, an oversampling technique to fix data imbalance, and evaluated on the defect prediction task.
Then, Agrawal et al. \cite{Agrawal2019Dodge} presented Dodge, hyperparameter optimization for machine learning.
Dodge detects and ignores redundant \emph{tunings} (i.e. configurations that lead to indistinguishable results) and runs orders of magnitude faster without harming the performance of the approaches under tuning. 
Dodge was evaluated in software defect prediction and text mining \cite{Agrawal2019Dodge}, later in bad smell detection, predicting GitHub issue close time, bug report analysis, and defect prediction~\cite{Agrawal2020HyperparameterOptimization}.
Yedida et Menzies \cite{Yedida2022Ghost} presented GHOST, a method that relies on a combination of hyper-parameter optimization of feedforward neural networks and a novel oversampling technique.
Lustosa et Menzies~\cite{lustosa2023optimizingSNEAK} present the hyperparameter optimization niSNEAK, based on landscape analytics, which was evaluated on the task of predicting the health of open source projects.
The implementations of these tools are tied to the optimization of machine learning models (e.g., Scikit-learn~\cite{scikit-learn} models). This prevents us from applying them in AST differencing.

To our knowledge, we are the first to propose and comprehensively study on hyperparameter optimization for AST differencing.

\vspace{-0.5cm}
\section{Conclusion}
\label{sec:conclusion}
In this paper, we have proposed to use hyperoptimization for AST differencing. We have described a novel approach, called \autotuningname, consisting of 
\begin{inparaenum}[\it 1)]
\item specifying the hyperparameter space of an AST differencing algorithm, 
\item applying a search technique to hyperoptimize the algorithm in a data-driven way based on a training dataset of AST differencing cases.
\end{inparaenum}
The approach has been instantiated for the popular AST differencing algorithm GumTree \cite{falleri_fine-grained_2014}.
We have performed a comprehensive quantitative assessment of \autotuningname, which shows that hyperoptimization improves the AST edit-scripts in up to 21.8\% of the differencing tasks using global hyperoptimization and up to 27\% using local hyperoptimization.
Our technique is widely applicable to all AST differencing algorithms: it can benefit both already proposed AST differencing systems and future ones to come. 
The main direction for future work is to apply and evaluate hyperparameter optimization of AST differencing  on other real-world scenarios, such as the improvement of downstream tasks using hyperoptimized AST edit-scripts (e.g., automated repair, bug fix analysis).

\section*{Acknowledgements}
This work was partially supported
by the Spanish Ministry of Science and Innovation (ref. RYC2021-031523-I), by the GAISSA Spanish research project (ref. TED2021-130923B-I00),
by the Wallenberg Artificial Intelligence, Autonomous Systems and Software Program (WASP) funded by Knut and Alice Wallenberg Foundation, and by the Swedish Foundation for Strategic Research (SSF). Some experiments were performed on resources provided by the Swedish National Infrastructure for Computing (SNIC).

\balance
\bibliographystyle{plain}
\bibliography{references}

\begin{thebibliography}{10}

\bibitem{Agrawal2019Dodge}
A.~{Agrawal}, W.~{Fu}, D.~{Chen}, X.~{Shen}, and T.~{Menzies}.
\newblock How to "dodge" complex software analytics.
\newblock {\em IEEE Transactions on Software Engineering}, pages 1--1, 2019.

\bibitem{Agrawal2018SMOTUNED}
Amritanshu Agrawal and Tim Menzies.
\newblock Is "better data" better than "better data miners"? on the benefits of
  tuning smote for defect prediction.
\newblock In {\em Proceedings of the 40th International Conference on Software
  Engineering}, ICSE '18, page 1050–1061. ACM, 2018.

\bibitem{Agrawal2020HyperparameterOptimization}
Amritanshu Agrawal, Xueqi Yang, Rishabh Agrawal, Rahul Yedida, Xipeng Shen, and
  Tim Menzies.
\newblock Simpler hyperparameter optimization for software analytics: Why, how,
  when?
\newblock {\em IEEE Transactions on Software Engineering}, 48(8):2939--2954,
  2022.

\bibitem{optuna_2019}
Takuya Akiba, Shotaro Sano, Toshihiko Yanase, Takeru Ohta, and Masanori Koyama.
\newblock Optuna: A next-generation hyperparameter optimization framework.
\newblock In {\em Proceedings of the 25th {ACM} {SIGKDD} International
  Conference on Knowledge Discovery and Data Mining}, 2019.

\bibitem{Apel2012MegeAutotuning}
Sven Apel, Olaf Le\ss{}enich, and Christian Lengauer.
\newblock Structured merge with auto-tuning: Balancing precision and
  performance.
\newblock In {\em Proceedings of the 27th IEEE/ACM International Conference on
  Automated Software Engineering}, ASE 2012, page 120–129. ACM, 2012.

\bibitem{Arcuri2013ParametertuningSE}
Andrea Arcuri and Gordon Fraser.
\newblock Parameter tuning or default values? an empirical investigation in
  search-based software engineering.
\newblock {\em Empirical Software Engineering}, 18, 06 2013.

\bibitem{Basios2018Darwinian}
Michail Basios, Lingbo Li, Fan Wu, Leslie Kanthan, and Earl~T. Barr.
\newblock Darwinian data structure selection.
\newblock In {\em Proceedings of the 2018 26th ACM Joint Meeting on European
  Software Engineering Conference and Symposium on the Foundations of Software
  Engineering}, ESEC/FSE 2018, page 118–128. ACM, 2018.

\bibitem{Bergstra2011AlgorithmsHyper}
James Bergstra, R\'{e}mi Bardenet, Yoshua Bengio, and Bal\'{a}zs K\'{e}gl.
\newblock Algorithms for hyper-parameter optimization.
\newblock In {\em Proceedings of the 24th International Conference on Neural
  Information Processing Systems}, NIPS’11, page 2546–2554. Curran
  Associates Inc., 2011.

\bibitem{Bergstra2012RandonHyper}
James Bergstra and Yoshua Bengio.
\newblock Random search for hyper-parameter optimization.
\newblock {\em J. Mach. Learn. Res.}, 13:281–305, February 2012.

\bibitem{Bergstra2015Hyperop}
James Bergstra, Brent Komer, Chris Eliasmith, Dan Yamins, and David Cox.
\newblock Hyperopt: A python library for model selection and hyperparameter
  optimization.
\newblock {\em Computational Science \& Discovery}, 8:014008, 07 2015.

\bibitem{Chawathe1996}
Sudarshan~S. Chawathe, Anand Rajaraman, Hector Garcia-Molina, and Jennifer
  Widom.
\newblock Change detection in hierarchically structured information.
\newblock {\em SIGMOD Rec.}, 25(2):493–504, June 1996.

\bibitem{10.1145/233269.233366}
Sudarshan~S. Chawathe, Anand Rajaraman, Hector Garcia-Molina, and Jennifer
  Widom.
\newblock Change detection in hierarchically structured information.
\newblock In {\em Proceedings of the 1996 ACM SIGMOD International Conference
  on Management of Data}, SIGMOD ’96, page 493–504. ACM, 1996.

\bibitem{Chawla2002SMOTE}
Nitesh~V. Chawla, Kevin~W. Bowyer, Lawrence~O. Hall, and W.~Philip Kegelmeyer.
\newblock Smote: Synthetic minority over-sampling technique.
\newblock {\em J. Artif. Int. Res.}, 16(1):321–357, jun 2002.

\bibitem{delaTorre2018Imprecision}
Guillermo de~la Torre, Romain Robbes, and Alexandre Bergel.
\newblock Imprecisions diagnostic in source code deltas.
\newblock In {\em Proceedings of the International Conference on Mining
  Software Repositories (MSR)}, page 492–502. ACM, 2018.

\bibitem{decker_srcdiff_2020}
Michael~John Decker, Michael~L Collard, L~Gwenn Volkert, and Jonathan~I
  Maletic.
\newblock {srcDiff}: {A} syntactic differencing approach to improve the
  understandability of deltas.
\newblock {\em Journal of Software: Evolution and Process}, 32(4):e2226, 2020.
\newblock Wiley Online Library.

\bibitem{dotzler_move-optimized_2016}
Georg Dotzler and Michael Philippsen.
\newblock Move-optimized source code tree differencing.
\newblock In {\em 2016 31st {IEEE}/{ACM} {International} {Conference} on
  {Automated} {Software} {Engineering} ({ASE})}, pages 660--671. IEEE, 2016.

\bibitem{falleri_fine-grained_2014}
Jean-Rémy Falleri, Floréal Morandat, Xavier Blanc, Matias Martinez, and
  Martin Monperrus.
\newblock Fine-grained and accurate source code differencing.
\newblock In {\em Proceedings of the 29th {ACM}/{IEEE} international conference
  on {Automated} software engineering}, pages 313--324, 2014.

\bibitem{fan2021differential}
Yuanrui Fan, Xin Xia, David Lo, Ahmed~E. Hassan, Yuan Wang, and Shanping Li.
\newblock A differential testing approach for evaluating abstract syntax tree
  mapping algorithms.
\newblock In {\em 2021 IEEE/ACM 43rd International Conference on Software
  Engineering (ICSE)}, pages 1174--1185, 2021.

\bibitem{Feurer2015AutoML}
Matthias Feurer, Aaron Klein, Katharina Eggensperger, Jost Springenberg, Manuel
  Blum, and Frank Hutter.
\newblock Efficient and robust automated machine learning.
\newblock In C.~Cortes, N.~D. Lawrence, D.~D. Lee, M.~Sugiyama, and R.~Garnett,
  editors, {\em Advances in Neural Information Processing Systems}, pages
  2962--2970. 2015.

\bibitem{Fluri2007Evolution}
B.~{Fluri}, M.~{Wursch}, and H.~C. {Gall}.
\newblock Do code and comments co-evolve? on the relation between source code
  and comment changes.
\newblock In {\em Working Conference on Reverse Engineering (WCRE)}, pages
  70--79, 2007.

\bibitem{fluri_change_2007}
Beat Fluri, Michael Wuersch, Martin Pinzger, and Harald Gall.
\newblock Change distilling: {Tree} differencing for fine-grained source code
  change extraction.
\newblock {\em IEEE Transactions on software engineering}, 33(11):725--743,
  2007.
\newblock IEEE.

\bibitem{Fraser2011Evosuite}
Gordon Fraser and Andrea Arcuri.
\newblock Evosuite: Automatic test suite generation for object-oriented
  software.
\newblock In {\em Proceedings of Foundations of Software Engineering}, ESEC/FSE
  '11, page 416–419. ACM, 2011.

\bibitem{frick_generating_2018}
Veit Frick, Thomas Grassauer, Fabian Beck, and Martin Pinzger.
\newblock Generating accurate and compact edit scripts using tree differencing.
\newblock In {\em {International} {Conference} on {Software} {Maintenance} and
  {Evolution} ({ICSME})}, pages 264--274. IEEE, 2018.

\bibitem{hashimoto_diffts_2008}
Masatomo Hashimoto and Akira Mori.
\newblock Diff/{TS}: {A} tool for fine-grained structural change analysis.
\newblock In {\em 2008 15th working conference on reverse engineering}, pages
  279--288. IEEE, 2008.

\bibitem{higo_generating_2017}
Yoshiki Higo, Akio Ohtani, and Shinji Kusumoto.
\newblock Generating simpler ast edit scripts by considering copy-and-paste.
\newblock In {\em 2017 32nd {IEEE}/{ACM} {International} {Conference} on
  {Automated} {Software} {Engineering} ({ASE})}, pages 532--542. IEEE, 2017.

\bibitem{huang_cldiff_2018}
Kaifeng Huang, Bihuan Chen, Xin Peng, Daihong Zhou, Ying Wang, Yang Liu, and
  Wenyun Zhao.
\newblock Cldiff: generating concise linked code differences.
\newblock In {\em Proceedings of the 33rd {ACM}/{IEEE} {International}
  {Conference} on {Automated} {Software} {Engineering}}, pages 679--690, 2018.

\bibitem{Hutter2011smbo}
Frank Hutter, Holger~H. Hoos, and Kevin Leyton-Brown.
\newblock Sequential model-based optimization for general algorithm
  configuration.
\newblock In {\em Proceedings of the 5th International Conference on Learning
  and Intelligent Optimization}, LION’05, page 507–523, Berlin, Heidelberg,
  2011. Springer-Verlag.

\bibitem{Jia2015Learing}
Y.~{Jia}, M.~B. {Cohen}, M.~{Harman}, and J.~{Petke}.
\newblock Learning combinatorial interaction test generation strategies using
  hyperheuristic search.
\newblock In {\em 2015 IEEE/ACM 37th IEEE International Conference on Software
  Engineering}, volume~1, pages 540--550, 2015.

\bibitem{kim2011evolutionapi}
M.~{Kim}, D.~{Cai}, and S.~{Kim}.
\newblock An empirical investigation into the role of api-level refactorings
  during software evolution.
\newblock In {\em 2011 33rd International Conference on Software Engineering
  (ICSE)}, pages 151--160, 2011.

\bibitem{Kotelyanskii2014}
Anton Kotelyanskii and Gregory~M. Kapfhammer.
\newblock Parameter tuning for search-based test-data generation revisited:
  Support for previous results.
\newblock In {\em 2014 14th International Conference on Quality Software},
  pages 79--84, 2014.

\bibitem{Kotthoff2016Autoweka}
Lars Kotthoff, Chris Thornton, Holger~H. Hoos, Frank Hutter, and Kevin
  Leyton-Brown.
\newblock Auto-weka 2.0: Automatic model selection and hyperparameter
  optimization in weka.
\newblock {\em Journal of Machine Learning Research}, 18(25):1--5, 2017.

\bibitem{fixminer}
Anil Koyuncu, Kui Liu, Tegawendé~F. Bissyandé, Dongsun Kim, Jacques Klein,
  Martin Monperrus, and Yves~Le Traon.
\newblock Fixminer: Mining relevant fix patterns for automated program repair.
\newblock {\em Empirical Software Engineering Journal, Springer Verlag}, 2020.

\bibitem{Le2016HDRepair}
X.~B.~D. Le, D.~Lo, and C.~L. Goues.
\newblock {History Driven Program Repair}.
\newblock In {\em Proceedings of the 23rd International Conference on Software
  Analysis, Evolution, and Reengineering (SANER)}, pages 213--224, 2016.

\bibitem{le2017s3}
Xuan-Bach~D. Le, Duc-Hiep Chu, David Lo, Claire Le~Goues, and Willem Visser.
\newblock S3: Syntax- and semantic-guided repair synthesis via programming by
  examples.
\newblock In {\em Proceedings of the 2017 11th Joint Meeting on Foundations of
  Software Engineering}, ESEC/FSE 2017, page 593–604. ACM, 2017.

\bibitem{Li2020CPDP}
Ke~Li, Zilin Xiang, Tao Chen, Shuo Wang, and Kay~Chen Tan.
\newblock Understanding the automated parameter optimization on transfer
  learning for {CPDP:} an empirical study.
\newblock {\em CoRR}, abs/2002.03148, 2020.

\bibitem{lustosa2023optimizingSNEAK}
Andre Lustosa and Tim Menzies.
\newblock Optimizing predictions for very small data sets: a case study on
  open-source project health prediction.
\newblock arXiv:2301.06577, 2023.

\bibitem{Martinez2013Models}
Matias Martinez and Martin Monperrus.
\newblock Mining software repair models for reasoning on the search space of
  automated program fixing.
\newblock {\em Empirical Software Engineering}, 20(1):176–205, February 2015.

\bibitem{matsumoto_beyond_2019}
Junnosuke Matsumoto, Yoshiki Higo, and Shinji Kusumoto.
\newblock Beyond {GumTree}: a hybrid approach to generate edit scripts.
\newblock In {\em 2019 {IEEE}/{ACM} 16th {International} {Conference} on
  {Mining} {Software} {Repositories} ({MSR})}, pages 550--554. IEEE, 2019.

\bibitem{Meng2013Lase}
N.~{Meng}, M.~{Kim}, and K.~S. {McKinley}.
\newblock Lase: Locating and applying systematic edits by learning from
  examples.
\newblock In {\em 2013 35th International Conference on Software Engineering
  (ICSE)}, pages 502--511, 2013.

\bibitem{Meng2011Sysedit}
Na~Meng, Miryung Kim, and Kathryn~S. McKinley.
\newblock Systematic editing: Generating program transformations from an
  example.
\newblock In {\em Proceedings of the Conference on Programming Language Design
  and Implementation}, page 329–342. ACM, 2011.

\bibitem{cvs-vintage}
Martin Monperrus and Matias Martinez.
\newblock {Conservation and Replication with CVS-Vintage: A Dataset of CVS
  Repositories of Java Software}.
\newblock Technical report, 2012.

\bibitem{myers1986ano}
Eugene~W Myers.
\newblock Ano (nd) difference algorithm and its variations.
\newblock {\em Algorithmica}, 1(1-4):251--266, 1986.

\bibitem{Nguyen2016}
Anh~Tuan Nguyen, Michael Hilton, Mihai Codoban, Hoan~Anh Nguyen, Lily Mast, Eli
  Rademacher, Tien~N. Nguyen, and Danny Dig.
\newblock Api code recommendation using statistical learning from fine-grained
  changes.
\newblock In {\em Proceedings of the 2016 24th ACM SIGSOFT International
  Symposium on Foundations of Software Engineering}, FSE 2016, page 511–522.
  ACM, 2016.

\bibitem{PANICHELLA20211LDAHyper}
Annibale Panichella.
\newblock A systematic comparison of search-based approaches for lda
  hyperparameter tuning.
\newblock {\em Information and Software Technology}, 130:106411, 2021.

\bibitem{pawlak:hal-01078532}
Renaud Pawlak, Martin Monperrus, Nicolas Petitprez, Carlos Noguera, and Lionel
  Seinturier.
\newblock {Spoon: A Library for Implementing Analyses and Transformations of
  Java Source Code}.
\newblock {\em {Software: Practice and Experience}}, 46:1155--1179, 2015.
\newblock update for oadoi on Nov 02 2018.

\bibitem{scikit-learn}
F.~Pedregosa, G.~Varoquaux, A.~Gramfort, V.~Michel, B.~Thirion, O.~Grisel,
  M.~Blondel, P.~Prettenhofer, R.~Weiss, V.~Dubourg, J.~Vanderplas, A.~Passos,
  D.~Cournapeau, M.~Brucher, M.~Perrot, and E.~Duchesnay.
\newblock Scikit-learn: Machine learning in {P}ython.
\newblock {\em Journal of Machine Learning Research}, 12:2825--2830, 2011.

\bibitem{Shahriari2015}
B.~{Shahriari}, K.~{Swersky}, Z.~{Wang}, R.~P. {Adams}, and N.~{de Freitas}.
\newblock Taking the human out of the loop: A review of bayesian optimization.
\newblock {\em Proceedings of the IEEE}, 104(1):148--175, 2016.

\bibitem{Shu2021SecurityReports}
Rui Shu, Tianpei Xia, Jianfeng Chen, Laurie Williams, and Tim Menzies.
\newblock How to better distinguish security bug reports (using dual
  hyperparameter optimization).
\newblock {\em Empirical Software Engineering}, 26(3):53, 2021.

\bibitem{Sobreira2018}
V.~{Sobreira}, T.~{Durieux}, F.~{Madeiral}, M.~{Monperrus}, and M.~{de Almeida
  Maia}.
\newblock Dissection of a bug dataset: Anatomy of 395 patches from defects4j.
\newblock In {\em 2018 IEEE 25th International Conference on Software Analysis,
  Evolution and Reengineering (SANER)}, pages 130--140, 2018.

\bibitem{Tantithamthavorn2019ImpactOptimization}
C.~{Tantithamthavorn}, S.~{McIntosh}, A.~E. {Hassan}, and K.~{Matsumoto}.
\newblock The impact of automated parameter optimization on defect prediction
  models.
\newblock {\em IEEE Transactions on Software Engineering}, 45(7):683--711,
  2019.

\bibitem{Tsantalis2018Refactor}
N.~{Tsantalis}, M.~{Mansouri}, L.~{Eshkevari}, D.~{Mazinanian}, and D.~{Dig}.
\newblock Accurate and efficient refactoring detection in commit history.
\newblock In {\em 2018 IEEE/ACM 40th International Conference on Software
  Engineering (ICSE)}, pages 483--494, 2018.

\bibitem{Villalobos2021}
Leonardo Villalobos-Arias and Christian Quesada-L\'{o}pez.
\newblock Comparative study of random search hyper-parameter tuning for
  software effort estimation.
\newblock In {\em Proceedings of the 17th International Conference on
  Predictive Models and Data Analytics in Software Engineering}, PROMISE 2021,
  page 21–29. ACM, 2021.

\bibitem{Yang2019Pruning}
C.~{Yang} and E.~J. {Whitehead}.
\newblock Pruning the ast with hunks to speed up tree differencing.
\newblock In {\em 2019 IEEE 26th International Conference on Software Analysis,
  Evolution and Reengineering (SANER)}, pages 15--25, 2019.

\bibitem{Yedida2022Ghost}
Rahul Yedida and Tim Menzies.
\newblock On the value of oversampling for deep learning in software defect
  prediction.
\newblock {\em IEEE Trans. Softw. Eng.}, 48(8):3103–3116, aug 2022.

\bibitem{Zhong2015Fixes}
H.~{Zhong} and Z.~{Su}.
\newblock An empirical study on real bug fixes.
\newblock In {\em 2015 IEEE/ACM 37th IEEE International Conference on Software
  Engineering}, 2015.

\end{thebibliography}

\section{Biography Section}

\begin{IEEEbiography}[{\includegraphics[width=1in,height=1.25in,clip,keepaspectratio]{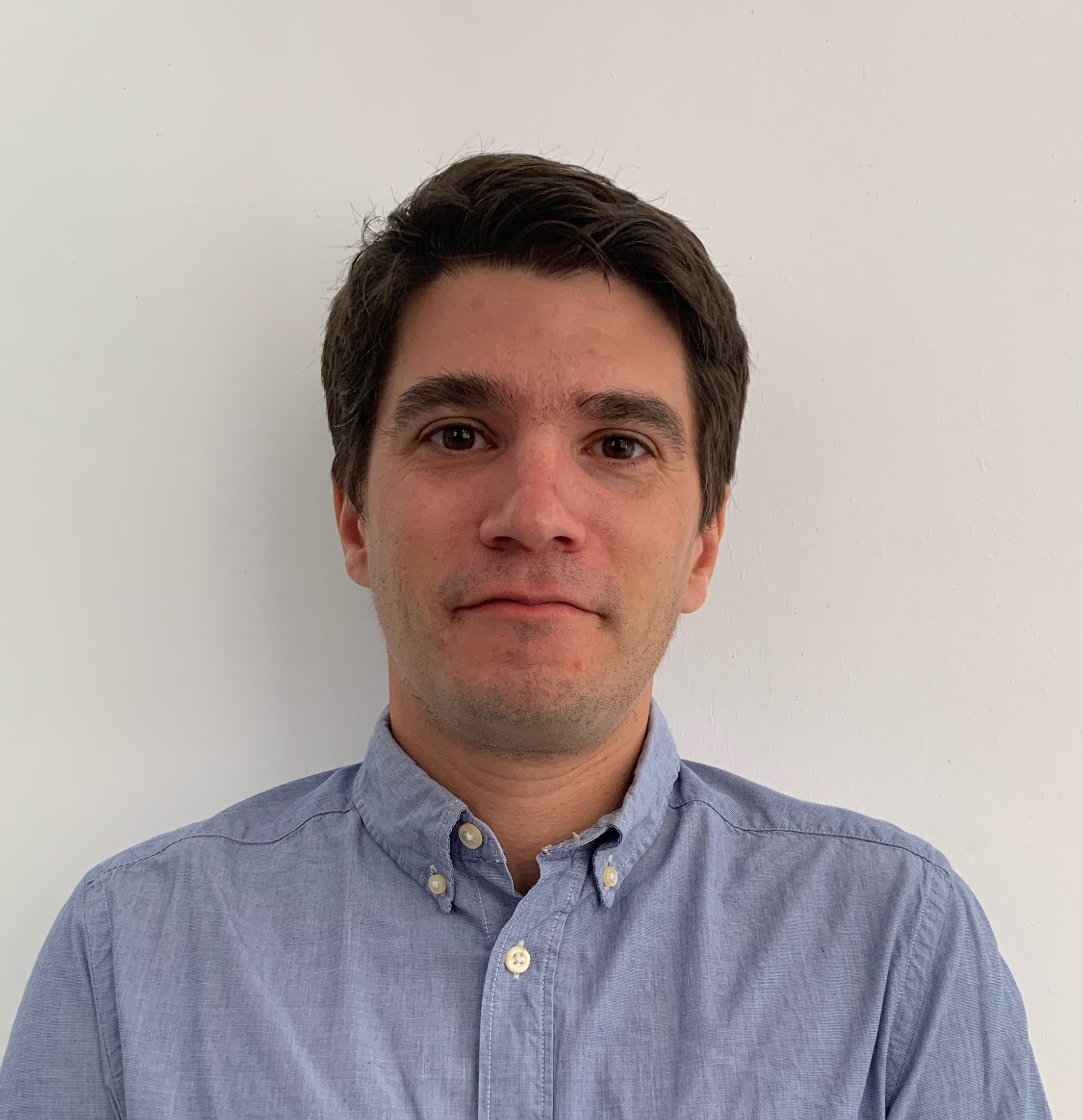}}]{Matias Martinez} is a researcher from the Universidad Politécnica de
Cataluña -BarcelonaTech- (Spain). He got his PhD degree from University of Lille (France) and a Computer Science degree from UNICEN
(Argentina).  
He was previously an associate professor at the Université Polytechnique Hauts-de-France (France). 
His research focuses on empirical software engineering,  automated software engineering, and software sustainability.
\end{IEEEbiography}

\begin{IEEEbiography}[{\includegraphics[width=1in,height=1.25in,clip,keepaspectratio]{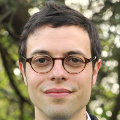}}]{Jean-Remy Falleri} is a full professor at Bordeaux INP and a researcher at the LaBRI laboratory. He is also a junior member at the "Institut Universitaire de France". He holds a computer science "Habilitation" degree from the University of Bordeaux and a Ph.D. degree from the University of Montpellier. His research interests revolve around software engineering and software evolution.
\end{IEEEbiography}

\begin{IEEEbiography}[{\includegraphics[width=1in,height=1.25in,clip,keepaspectratio]{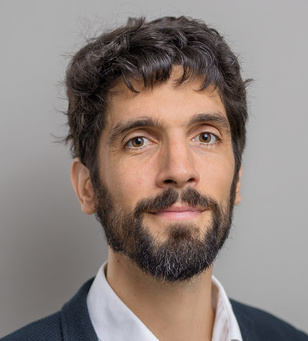}}]{Martin Monperrus} is Professor of Software Technology at KTH Royal Institute of Technology. His research lies in the field of software engineering with a current focus on automatic program repair, AI on code and program hardening. He received a Ph.D. from the University of Rennes, and a Master's degree from Compiègne University of Technology. Homepage: https://www.monperrus.net/martin/
\end{IEEEbiography}

\end{document}